\documentclass[10pt]{iopart}
\usepackage{url} 
\usepackage{pdflscape} 
\usepackage{amsmath}
\usepackage{amssymb} 
\usepackage{epsfig} 
\usepackage{amsfonts}
\usepackage{graphicx}
\usepackage{subfigure}
\usepackage{graphicx}
\usepackage{times,fancyhdr}

\usepackage{enumitem}
\setlist[enumerate,2]{label=\roman*)}

\def\case#1/#2{\textstyle\frac{#1}{#2}}

\newcommand{\be}{\begin{equation}}
\newcommand{\ee}{\end{equation}}
\newcommand{\ben}{\begin{eqnarray}}
\newcommand{\een}{\end{eqnarray}}
\usepackage{mdframed}
\usepackage{grffile}
\usepackage{amssymb}
\usepackage{ulem}
\usepackage[T1]{fontenc}
\usepackage[colorlinks=true,
            linkcolor=blue,
           urlcolor=blue,
           citecolor=blue]{hyperref}
\usepackage{subfigure}
\newtheorem{thm}{Theorem}[section]
\newtheorem{cor}{Corollary}[thm]

\providecommand{\U}[1]{\protect\rule{.1in}{.1in}}
\newcommand{\mincir}{\raise
-3.truept\hbox{\rlap{\hbox{$\sim$}}\raise4.truept\hbox{$<$}\ }}
\newcommand{\magcir}{\raise
-3.truept\hbox{\rlap{\hbox{$\sim$}}\raise4.truept\hbox{$>$}\ }}

\def\S{{}^3\!S_+}

\begin{document}

\title{Generalized scalar field cosmologies: theorems on asymptotic behavior}

\author{Genly Leon}
\address{Departamento  de  Matem\'aticas,  Universidad  Cat\'olica  del  Norte, Avda. Angamos  0610,  Casilla  1280  Antofagasta,  Chile.}
\ead{genly.leon@ucn.cl}

\author{Felipe Orlando Franz Silva}
\address{Departamento  de  Matem\'aticas,  Universidad  Cat\'olica  del  Norte, Avda. Angamos  0610,  Casilla  1280  Antofagasta,  Chile.}
\ead{felipe.franz@alumnos.ucn.cl}

\begin{abstract}
Phase-space descriptions are used to find qualitative features of the solutions
of generalized scalar field cosmologies with arbitrary potentials and arbitrary
couplings to matter. Previous results are summarized and new ones are presented 
as theorems, which include the previous ones as corollaries. Examples
of these results are presented as well as counterexamples when the hypotheses
of the theorems are not fulfilled. Potentials with small cosine-like corrections
motivated by inflationary loop-quantum cosmology are discussed. Finally, the
Hubble-normalized formulation for the FRW metric and for the Bianchi I metric 
is applied to a scalar field cosmology with a generalized harmonic potential,
non-minimally coupled to matter, and the stability of the solutions is discussed.
\end{abstract}


\maketitle

\section{Introduction}

Scalar fields are used to describe the gravitational field. Some scalar field theories of special interest are the scalar-tensor theories such as Jordan theory, \cite{Jordan:1958zz} as a generalization of the Kaluza-Klein Theory, the Brans-Dicke Theory \cite{Brans:1961sx}, Horndeski Theories \cite{Horndeski:1974wa}, Teleparalell Analogue of Horndeski Theories \cite{Bahamonde:2019shr,Bahamonde:2019ipm,Bahamonde:2020cfv}, Inflationary Models \cite{Guth:1980zm}, Extended Quintessence, Modified Gravity, Ho\v{r}ava-Lifschitz and the Galileons, etc.,  \cite{Copeland:1993jj,Ibanez:1995zs,Chimento:1995da,Lidsey:1995np,Coley:1997nk,Copeland:1998fz,Coley:1999mj,Coley:1999uh,Coley:2000zw,Coley:2000yc,Coley:2003tf,Elizalde:2004mq,Capozziello:2005tf,Curbelo:2005dh,Gonzalez:2005ie,Gonzalez:2006cj,Lazkoz:2006pa,Lazkoz:2007mx,Elizalde:2008yf,Leon:2009dt,Leon:2009rc,Leon:2009ce,Leon:2010pu,Basilakos:2011rx,Xu:2012jf,Leon:2012mt,Leon:2013qh,Fadragas:2013ina,Kofinas:2014aka,Leon:2014yua,Paliathanasis:2014yfa,DeArcia:2015ztd,Paliathanasis:2015gga,Leon:2015via,Barrow:2016qkh,Barrow:2016wiy,Cruz:2017ecg,Paliathanasis:2017ocj,Alhulaimi:2017ocb,Dimakis:2017kwx,Giacomini:2017yuk,Karpathopoulos:2017arc,DeArcia:2018pjp,Tsamparlis:2018nyo,Paliathanasis:2018vru,Basilakos:2019dof,VanDenHoogen:2018anx,Leon:2018lnd,Leon:2018skk,Leon:2019mbo,Paliathanasis:2019qch,Leon:2019jnu,Paliathanasis:2019pcl,Barrow:2018zav,Quiros:2019ktw,Barker:2020elg,Leon:2019iwj,Giacomini:2020zmv,Paliathanasis:2020abu,Giacomini:2020grc}.

There are several studies in literature that provide both global and local dynamical systems analysis for scalar field cosmologies with arbitrary potentials and arbitrary couplings. 
In \cite{Foster:1998sk}, a very large and natural class of scalar field models having an arbitrary non-negative potential function $V(\phi)$ with a flat Friedmann-Lema\^{i}tre-Robertson-Walker (FLRW) metric was studied; yielding to a simple and regular past asymptotic structure which corresponds to the exact integrable massless scalar field cosmologies with the exception of a set that has zero measure.  This model was generalized in \cite{Miritzis:2003ym} for flat and negatively curved FLRW models by adding a perfect fluid matter source to the scalar field. In particular, for  a scalar field with  potential with a local zero minimum, the universe ever expands and the energy density asymptotically approaches zero. Additionally, the scalar field asymptotically reaches the minimum of the potential. On the other hand, a closed FLRW model with ordinary matter can avoid a re-collapse due to the presence of a scalar field with a non-negative potential.

The model in \cite{Miritzis:2003ym} was extended in \cite{Dania&Yunelsy,Leon:2008de} to a scalar field non-minimally coupled to matter. This scenario incidentally contains a particular realization of the model in \cite{Giambo:2009byn}, which arises in the conformal frame of $F(R)$ theories non-minimally coupled to matter. It was proved that,
under generic hypotheses, the future attractor corresponds to the vacuum de Sitter solution by considering an arbitrary potential  $V(\phi)$ and arbitrary coupling function $\chi(\phi)$. Also, the fact that the scalar field diverges into the past was proved in \cite{Dania&Yunelsy,Leon:2008de};  extending the previous results in \cite{Foster:1998sk,Miritzis:2003ym}. In order to study the dynamics close to the initial singularity, the limit $\phi\rightarrow \infty$ was considered by imposing some regularity conditions on the potential and on the coupling function.  Interestingly, the asymptotic structure of solutions towards the past was proved to be  simple and regular, and independent of the features of the potential, the coupling function and the background matter as in \cite{Miritzis:2003ym}. The dynamics of a non-minimally coupled scalar field model with a $(1-\xi \phi^2) R$ coupling and potentials $V(\phi)=V_0(1+\phi^p)^2$ and $V(\phi)=V_0e^{\lambda \phi^2}$ were presented in \cite{Shahalam:2019jgs}. Other non-minimally coupled scalar field models were studied in e.g.:  \cite{Nojiri:2019riz,Humieja:2019ywy,Matsumoto:2017gnx,Matsumoto:2015hua,Solomon:2015hja,Harko:2015pma,Minazzoli:2014xua,Skugoreva:2013ooa,Jamil:2012vb,Miritzis:2011zz,Hrycyna:2007gd}.

In reference \cite{Giambo:2008ck}, homogeneous FLRW cosmological models with a self--interacting scalar field source were studied  for the flat, negatively and positively curved models. The analysis incorporates a wide class of self--interacting potentials, and only requires a scalar field potential to be bounded from below and divergent when the field diverges, say, positive potentials which exhibit asymptotically polynomial or exponential behaviors. Potentials with a negative inferior bound lead asymptotically to Anti de Sitter (AdS) solutions.

In reference \cite{Giambo:2009byn}, the evolution of a cosmological model with a perfect fluid matter source with energy density $\rho_m$ and pressure $p_m$, satisfying the equation of state $p_m=(\gamma-1) \rho_m$ was studied. A scalar field with self--interacting potential $V(\phi)$ non-minimally coupled to matter with an exponential coupling $\chi(\phi)$ (in the sense of \cite{Dania&Yunelsy,Leon:2008de}) was added, and flat and negatively curved FLRW models were considered. The existence of a very generic class of potentials having an equilibrium point which corresponds to the non-negative local minimum for $V(\phi)$, which is asymptotically stable was proved in \cite{Giambo:2009byn}. The same happens for horizontal asymptotes that are approached from above by $V(\phi)$. Furthermore, in this reference, all flat models were classified for which one of the matter constituents will eventually dominate. Particularly, if the barotropic matter index $\gamma$ is larger than 1, generically, there is an energy transfer from the fluid to the scalar field which eventually dominates over the background matter.

The original models in \cite{Foster:1998sk,Miritzis:2003ym} were gradually extended to more general scenarios in \cite{Leon:2010ai,Leon:2014bta,Leon:2014rra,Fadragas:2014mra}. In \cite{Fadragas:2014mra}, the flat FLRW model in the conformal Einstein's frame of scalar-tensor gravity theories for arbitrary positive potentials and arbitrary coupling functions was studied. Radiation was incorporated into the matter content to obtain a more realistic scenario. In \cite{Leon:2010ai,Fadragas:2014mra}, a procedure for the analysis in the limit $\phi \rightarrow \infty$ was implemented by using a suitable change of variables. The method has been exemplified for: (a) a non--minimally coupled scalar field model with a double exponential potential $V(\phi)= V_1 e^{\alpha \phi}+ V_2 e^{\beta \phi}$, $\alpha$ and $\beta$ are constants that satisfy $0<\alpha<\beta$, and
a coupling function $\chi=\chi_0 e^{\frac{\lambda \phi}{4-3\gamma}}$ where $\lambda$ is a constant (discussed in \cite{Tzanni:2014eja}); 
(b) a non--minimally coupled scalar field model with the Albrecht-Skordis potential $V(\phi)= e^{-\mu \phi}\left(A-(\phi-B)^2\right)$ and a power-law coupling $\chi(\phi)=\left(\frac{3\alpha}{8}\right)^{\frac{1}{\alpha}}\chi_0 \left(\phi-\phi_0\right)^{\frac{2}{\alpha}}$ where $\alpha>0$ and $\phi_0\geq 0$ are constant, originally investigated in \cite{Leon:2008de} for a less general model, which contains the cases investigated in \cite{vandenHoogen:1999qq,Albrecht:1999rm,Copeland:1997et} as particular cases.

In \cite{Tzanni:2014eja}, a flat FLRW model with a perfect fluid source and a scalar field with double exponential potential which is non-minimally coupled to matter was studied. The coupling is derived from the formulation of the $F(R)$- gravity as an equivalent scalar-tensor theory. Conditions were provided for which $\rho_m\rightarrow 0, \dot \phi\rightarrow 0$ and $\phi \rightarrow +\infty$ as $t\rightarrow \infty$ (see  Proposition 1 of \cite{Tzanni:2014eja}), and conditions under which $H$ and $\phi$ blow-up in a finite time  (see  Proposition 2 of \cite{Tzanni:2014eja}).  In the reference \cite{Giambo:2019ymx} was studied the late time evolution of a negatively curved FLRW model  with a perfect fluid matter source  with energy density $\rho_m$ and pressure  $p_m=(\gamma-1) \rho_m$, and a scalar field non-minimally coupled to matter. Under mild assumptions on the potential, it was found that non-negative local minima of $V$ are asymptotically stable. For non-degenerated minima with zero critical value, it was proved that for $\gamma> 2/3$, there is a transfer of energy from the fluid and the scalar field to the energy density of the scalar curvature, in contrast to the previous bound $\gamma> 1$ for a flat FLRW model. Thus, if there is a scalar curvature,
it has a dominant effect on the late evolution of the universe, and will eventually dominate both the perfect fluid and the scalar field. The analysis was complemented with the case where $V$ is an exponential potential; therefore, the scalar field diverges to infinity.

In \cite{Cid:2017wtf} was presented  a generalized Brans-Dicke Lagrangian including a non-minimally coupled Gauss-Bonnet term without imposing the vanishing torsion condition.  The existence of exact solutions and integrable dynamical systems in  Chiral cosmology (wherein  non-linear expressions of the kinetic term of the scalar fields exist) were further studied, and their  cosmological consequences examined in \cite{Paliathanasis:2018vru}. Some exact analytic solutions for a system of $N$-scalar fields were presented. Furthermore, some studies of cosmological effects of scalar fields and their effects in multiple-field
inflation are: \cite{Khlopov:1985jw,Sakharov:1993qh,Sakharov:1994pr,Rubin:2001yw,Khlopov:2002yi,Khlopov:2004sc,Khlopov:2008qy}. In \cite{Paliathanasis:2020abu} a detailed analysis for the asymptotic behaviour for the multi-scalar field Chiral
cosmological scenario was performed. It was proved that the maximum number of scalar fields which provides interesting physical results is $N=2$, while for $N > 2$ the new stationary points are only of mathematical interest since they do not describe new physics.

In the Einstein--Klein--Gordon system for a single-scalar field cosmology, the Raychaudhuri equation always decouples for a scalar field with exponential potential $V(\phi)= V_0 e^{-\lambda \phi}$. This is due to the fact it has the symmetry such that its derivative is also an exponential function.   The asymptotics of the remaining reduced system is then typically given by equilibrium points and often can be determined by a dynamical system analysis \cite{Coley:1999uh,Coley:2003mj,wainwrightellis1997}.  For other potentials that do not satisfy the above symmetry, like  the harmonic potential $V(\phi)= \mu^2 \phi^2$, the Hubble normalized equations  are augmented by the Raychaudhuri equation \cite{Alho:2014fha}. Furthermore, a local stability analysis is also difficult to apply due to the oscillations which are entering the system via Klein--Gordon equation \cite{Fajman:2020yjb}. Complementary  formulations  based  on  global variables  are  implemented  in  the companion paper \cite{PaperII}.

This paper is organized as follows: In Section \ref{Model}  a scalar field which has an arbitrary self--interacting potential is investigated. This scalar field is non-minimally coupled to matter through an arbitrary coupling function in which we analyze the corresponding cosmology. In Section \ref{SECT:Non_min} are presented the new Theorems \ref{Proposition I}, \ref{Theorem5.2}, \ref{Theorem5.3}, \ref{thmIIIFINAL}, and Corollaries \ref{Proposition II} and \ref{Proposition III} which are collectively referred as Leon \& Franz-Silva 2020. Well-known results like Corollaries \ref{PropositionIb}, \ref{tm}, \ref{PropositionII}, \ref{thm2.1}, \ref{Prop4Miritzis} and \ref{thm2.2} are recovered. Sections \ref{Sect2.4} and \ref{Sect.2.5} are  devoted to a dynamical system formulation for a scalar-field cosmology with generalized harmonic potentials $V_1(\phi)=\mu^3 \left[\frac{\phi^2}{\mu} + b f \cos\left(\delta + \frac{\phi}{f}\right)\right]$, $b\neq 0$ and $V_2(\phi)= \mu ^3 \left[b f \left(\cos (\delta )-\cos \left(\delta +\frac{\phi }{f}\right)\right)+\frac{\phi ^2}{\mu}\right]$, $b\neq 0$ in a vacuum respectively. These potentials incorporate cosine-like corrections with small phase motivated by inflationary loop-quantum cosmology \cite{Sharma:2018vnv}.  These potentials provide some examples and  counterexamples of the  theorems proved. In Section \ref{SECT:New4.5}  the Hubble-normalized formulation is used for the FLRW metric and for the Bianchi I metric for a scalar field cosmology with generalized harmonic potential $V(\phi)= \frac{\phi^2}{2} + f \left[1-\cos\left( \frac{\phi}{f}\right)\right]$, $f> 0$, non--minimally coupled to matter with coupling function  $\chi=\chi_0 e^{\frac{\lambda \phi}{4-3\gamma}}$ where  $\lambda$ and $\chi_0$ are constants and  $0\leq \gamma \leq 2, \quad \gamma \neq \frac{4}{3}$. 
In Section \ref{discussion} the main results are summarized and in Section \ref{Sect:7} conclusions are presented.

\section{Theorems on Asymptotic Behavior}
\label{Model}

The action for a general class of Scalar Tensor Theory of Gravity is written in the so-called Einstein's frame as \cite{Kaloper:1997sh,Gonzalez:2007ht}:
\begin{align}&
 {\cal S}=\int  d{ }^4 x \sqrt{|g|}\left[\frac{1}{2} R-\frac{1}{2} g^{\mu
\nu}\nabla_\mu\phi\nabla_\nu\phi-V(\phi)-\Lambda+\chi(\phi)^{-2}
\mathcal{L}(\mu,\nabla\mu,\chi(\phi)^{-1}g_{\alpha\beta})\right],\label{eq1}
\end{align} where a system of units in which $8\pi G=c=\hbar=1$ is used.
In equation \eqref{eq1}, $R$ is the curvature scalar, $\phi$ is the 
scalar field,  $\nabla_\alpha$
is the covariant derivative, $V(\phi)$ is the quintessence self-interaction potential, $\Lambda$ is the Cosmological Constant,
$\chi(\phi)^{-2}$ is the coupling function, $\mathcal{L}$
is the matter Lagrangian, and $\mu$ is the collective name for the
matter degrees of freedom. 

Additionally, it can be proved that the action
in the metric formalism defined by
\cite{Sotiriou:2008rp,DeFelice:2010aj}:
\begin{equation}
 {\cal S}_{\text{metric}}=\int  {d} x^4
\sqrt{|g|}\left[\frac{1}{2}F\left(  R\right)+{\cal
L}(\mu,\nabla\mu,g_{\alpha \beta})\right],\label{fRaction}
 \end{equation} 
where $F(R)$ is a function of the Ricci scalar $R$, and  ${\cal
L}$ accounts for the matter content of the universe  can be mapped under the conformal transformation $\widetilde{g}%
_{\mu\nu}=F^{\prime}\left(  R\right)  g_{\mu\nu}$ onto
\cite{Bean:2006up}:
\begin{equation}
\widetilde{S}=\int d^{4}x\sqrt{|\widetilde{g}|}\left\{\frac{1}{2}
\widetilde{R}-\frac{1}{2} \widetilde{g}^{\mu
\nu} {\nabla}_\mu\phi  {\nabla}_\nu\phi-V\left( \phi\right) 
 +e^{-2\sqrt{2/3}\phi}\mathcal{L}\left(\mu,\nabla\mu,
e^{-\sqrt{2/3}\phi
}\widetilde{g}\right)  \right\}, \label{action}%
\end{equation}
where a new scalar
field non--minimally coupled to matter
\begin{equation}
\phi=\sqrt{\frac{3}{2}}\ln F^{\prime}\left(  R\right), \label{scfi}%
\end{equation}
it appears in the theory. It is assumed that (\ref{scfi}) can be solved for $R$ to obtain a
function $R\left(  \phi\right),$ and the self--interacting potential of the scalar
field is given by
\begin{equation}
V\left(  \phi  \right)  =\frac{1}{2\left(
F^{\prime}\left(  R\left(  \phi\right)  \right)\right)
^{2}}\left(  R\left(  \phi\right) F^{\prime}\left(  R\left(  \phi\right)  \right)-F\left(  R\left(  \phi\right)  \right)\right). \label{pote}%
\end{equation}
The restrictions on the potential \eqref{pote} in the papers
\cite{Rendall:2004ic,Rendall:2005if,Rendall:2005fv,Rendall:2006cq}
were used in \cite{Macnay:2008nw} to impose conditions on the
function $F\left( R\right)$. It is easy to note that the model arising from the action
\eqref{fRaction} can be obtained from \eqref{eq1} by setting
$\chi(\phi)=e^{\sqrt{2/3}\phi}, \Lambda=0.$

For the action \eqref{eq1}, the matter energy-momentum tensor is defined by:
\begin{equation}T_{\alpha
\beta}=-\frac{2}{\sqrt{|g|}}\frac{\delta}{\delta g^{\alpha
\beta}}\left\{\sqrt{|g|}
 \chi^{-2}\mathcal{L}(\mu,\nabla\mu,\chi^{-1}g_{\alpha
 \beta})\right\}.\label{Tab}\end{equation}
The ``energy exchange'' vector is defined as: 
\begin{equation}
    Q_\beta\equiv\nabla^\alpha T_{\alpha \beta}=-\frac{1}{2}T\frac{1}{\chi(\phi)}\frac{\mathrm{d}\chi(\phi)}{\mathrm{d}\phi}\nabla_{\beta}\phi,\;
 T=T^\alpha_\alpha,
\end{equation}
 where $T$ is the trace of the energy-momentum tensor. 
 Additionally,  the geometric properties of the metric are incorporated in the form of the function: 
\begin{align}
&G_0(a)=\left\{ \begin{array}{cc}
-3\frac{k}{a^2}, k=0, \pm 1, & \text{FLRW}\\
\frac{\sigma_0^2}{a^6}, & \text{Bianchi I}
\end{array} 
\right.,
\end{align} to obtain the equations of motion for a  scalar field cosmology with the scalar field non-minimally coupled to matter, given by:
\begin{subequations}
	\label{Non_min2}
	\begin{align}
	& \dot{H}=-\frac{1}{2}\left(\gamma \rho_m+y^2\right)+\frac{1}{6}a G_0'(a), \label{Rachd}\\
	&\dot{\rho_m}=-3\gamma H\rho_m-\frac{1}{2}(4-3\gamma)\rho_m y\frac{d\ln\chi(\phi)}{d \phi}, \label{consb}\\
	& \dot a= a H,\\
	&\dot y=-3 H y -\frac{d V(\phi)}{d\phi}+\frac{1}{2}(4-3\gamma)\rho_m \frac{d\ln\chi(\phi)}{d \phi} \label{13EQ},\\
	& \dot\phi=y, \\
	& 3H^2=\rho_m+\frac{1}{2}\dot\phi^2+V(\phi)+\Lambda+G_0(a),\label{Fried2b}
	\end{align}
\end{subequations}
where  $\Lambda\geq 0$ is assumed, and  $\gamma \in [1,2]$.  Using equation \eqref{Fried2b}, the phase-space is defined as:
\begin{equation}
\label{Fried2bB}
   \left\{(H, \rho_m, a, y, \phi)\in \mathbb{R}^5: 3H^2=\rho_m+\frac{1}{2}y^2+V(\phi)+\Lambda+G_0(a)\right\}.
\end{equation}

\subsection{Main Theorems}
\label{SECT:Non_min}

Firstly, the cases  $G_0(a)=-3\frac{k}{a^2}, k=0, - 1$ and $G_0(a)=\frac{\sigma_0^2}{a^6}$ are studied, which are special cases of $G_0(a)= \frac{K^2}{a^p} \geq 0$. $K=0$ corresponds to flat FLRW metric; $K^2=1, p=2$ corresponds to negatively curved  FLRW metric; and  $K^2=\sigma_0^2, p=6$ corresponds to Bianchi I metric. The FLRW model  with positive curvature, $k=+1$, will be discussed in Section \ref{positive-k}.
The phase space $\Omega$ is defined as $\Omega=\left\{(H, \rho_m, a, y, \phi)\in \mathbb{R}^5: 3H^2=\rho_m+\frac{1}{2}y^2+V(\phi)+\Lambda+\frac{K^2}{a^p}, \rho_m\geq 0\right\}$ where it is assumed that $V(\phi)\geq 0$ is a function of class $C^2(\mathbb{R})$ and that $V(\phi)$ has a local minimum at $\phi=0$ and $V(0)=0$. In this case $(0,0,a_*,0,0)$, $a_*\rightarrow +\infty$ is an equilibrium configuration for the flow of \eqref{Non_min2}. The set  $\left\{(H, \rho_m, a, y, \phi)\in \Omega: H=0\right\}$ is invariant for the flow of  \eqref{Non_min2} and $H$ does not change sign.  On the contrary, if there is an orbit with  $H(0)>0$ and $H(t_1)<0$ for some $t_1>0$, this solution passes through the origin violating the existence and uniqueness of the solutions of a $C^1$ flow.

\begin{thm}[Leon \& Franz-Silva, 2020]
\label{Proposition I}
Assuming the following holds:
\begin{enumerate}
\item  $V(\phi)\in C^2(\mathbb{R}), V(\phi)\geq 0$, and $V(\phi)=0$ if and only if $\phi=0$. 
\item  $V^{\prime}(\phi)$ is bounded on $A\subset\mathbb{R}$ if $V(\phi)$ is bounded on  $A$.
\item $\Lambda\geq 0$, and $G_0(a)\geq 0$ has a negative power-law functional form $G_0(a)=\frac{K^2}{a^p}, p>0$. 
\item $\chi(\phi)\in C^2(\mathbb{R})$, and for all $A\subset \mathbb{R}$ there is a non-negative constant $K_1$, possibly depending on $A$, such as $\left| \frac{\chi '(\phi)}{\chi (\phi)}\right|\leq K_1$ for all $\phi \in A$. 
\end{enumerate}
Then, $\lim_{t\rightarrow \infty} \left( \rho_m, y, \frac{K^2}{a^p}\right)=(0,0,0)$.
\end{thm}
\textbf{Proof}.  Let $O^+(x_0)$ be the positive orbit that passes  through the regular point $x_0\in
\left\{(H, \rho_m, a, y, \phi)\in \Omega: H>0\right\}$ at the time $t_0$. Since $H$ is positive and decreases along $O^+(x_0)$, the limit $\lim_{t\rightarrow \infty} H(t)$ exists, and it is a non-negative number $\eta$. Furthermore, $H(t)\leq H(t_0)$ for all $t>t_0$. Then, 
 $\rho_m(t)+\frac{1}{2}y(t)^2+V(\phi(t))+\Lambda+\frac{K^2}{a(t)^p}= 3H(t)^2\leq 3 H(t_0)^2$ for all $t>t_0$. All above terms are non-negative,  so it follows that $\rho_m, \frac{1}{2}y(t)^2, \Lambda, \frac{K^2}{a(t)^p}$ are bounded by  $3 H(t_0)^2$  for all $t>t_0$.
Defining the set $A=\{\phi\in \mathbb{R}: V(\phi)\leq 3 H(t_0)^2\}$. Then, the orbit $O^+(x_0)$ is such that $\phi$  remains at the interior of $A$ for all $t>t_0$. 
 Given $G_0(a)=\frac{K^2}{a^p}, p>0$, the equation  \eqref{Rachd} can be written as:  
\begin{equation}
 \dot{H}=-\frac{1}{2}\left(\gamma \rho_m+y^2\right)-\frac{K^2 p}{6 a^p}.\label{Rachd2} 
\end{equation}
Then, by integration, it follows that: 
\begin{equation*}
H(t_0)-H(t)=\int_{t_0}^t \left(\frac{1}{6} K^2 p a(s)^{-p}+\frac{1}{2}
   \gamma  \rho_m (s)+\frac{1}{2} y(s)^2\right) \, ds.
\end{equation*}
Taking the limit as $t\rightarrow +\infty$, it is obtained 
\begin{equation*}
H(t_0)-\eta= H(t_0) -\lim_{t\rightarrow \infty} H(t)=\int_{t_0}^\infty \left(\frac{1}{6} K^2 p a(s)^{-p}+\frac{1}{2}
   \gamma  \rho_m (s)+\frac{1}{2} y(s)^2\right) \, ds.
\end{equation*}
From this equation, the improper integral is convergent:
\begin{equation*}
\int_{t_0}^\infty \left(\frac{1}{6} K^2 p a(s)^{-p}+\frac{1}{2}
   \gamma  \rho_m (s)+\frac{1}{2} y(s)^2\right) \, ds<\infty.
   \end{equation*}
Defining $f(t)=\left(\frac{1}{6} K^2 p a(t)^{-p}+\frac{1}{2}
   \gamma  \rho_m (t)+\frac{1}{2} y(t)^2\right)$, and taking the $t$ derivative, it follows\\
$\frac{d}{dt} f(t)  = -y V'(\phi)+H  \left(-\frac{1}{6} K^2 p^2 a^{-p}-\frac{3}{2} \left(\gamma ^2 \rho_m+2 y^2\right)\right)+\frac{(\gamma -2) (3 \gamma -4) \rho_m y 
   \chi '(\phi)}{4 \chi (\phi)}$. 
Using the results that $\rho_m, \frac{1}{2}y(t)^2$ and $\frac{K^2}{a(t)^p}$ are bounded by $3 H(t_0)^2$  for all $t>t_0$ and the hypothesis for $\chi$, it follows $\left|\frac{d}{dt} f(t)\right|\leq  y |V'(\phi)|+\frac{1}{6}p^2 H \left| K^2 a^{-p}\right|+\frac{3}{2}\gamma^2 \rho_m H +3 y^2 H+ \left|\frac{(\gamma -2) (3 \gamma -4)}{4}\right| \rho_m y \left|\frac{
   \chi '(\phi)}{ \chi (\phi)}\right|\leq \sqrt{6}  H(t_{0})
   \left|V'(\phi (t))\right| +\frac{1}{2}  H(t_{0})^3 (9\gamma^2 + p^2 +36)  +  \frac{3 \sqrt{6}}{4} H(t_{0})^3 K_1 |(\gamma -2) (3 \gamma -4)|$, 
for all $t>t_0$ along the positive orbit $O^+(x_0)$. Finally, given that $V(\phi)$ is bounded on $A$, $V^{\prime}(\phi)$ will be bounded on $A$ as well, leading to $\left|\frac{d}{dt} f(t)\right|<\infty$ through the positive orbit $O^+(x_0)$. Summarizing, the function $f(t)$ is non-negative, it has a bounded derivative through the orbit $O^+(x_0)$, and $\int_{t_0}^\infty f(s) \, ds$ is convergent.  Hence, 
  $\lim_{t\rightarrow \infty} f(t)=0$, from which, along with the non-negativeness of each term of $f(t)$,  we have $\lim_{t\rightarrow \infty} \left( \rho_m, y, \frac{K^2}{a^p}\right)=(0,0,0)$.  $\blacksquare$
	\\
	Now, it will be shown how Theorem \ref{Proposition I} generalizes previous results in the following corollaries: 
\begin{enumerate}
    \item[(A)] 
\begin{cor}[Leon \& Franz-Silva, 2020]\label{PropositionI}
Assuming that hypotheses i), ii) and iii) of Theorem \ref{Proposition I} are satisfied, and $\chi (\phi) \equiv 1$. Then  $\lim_{t\rightarrow \infty} \left( \rho_m, y, \frac{K^2}{a^p}\right)=(0,0,0)$.
\end{cor}
\textbf{Proof}.  Set  $\chi (\phi) \equiv 1$ (minimal coupling) in Theorem \ref{Proposition I}.

\item[(B)]  
\begin{cor}[Miritzis 2003. Proposition 2 of \cite{Miritzis:2003ym}] 
\label{PropositionIb}
Assuming that hypotheses i) and ii) of Theorem \ref{Proposition I} are satisfied, and $\chi (\phi) \equiv 1$, $\Lambda=0$,  $G_0(a)\equiv 0$, then $\displaystyle{\lim_{t\rightarrow  \infty} (\rho_m, y) =(0, 0)}$.
\end{cor}
\textbf{Proof}. Set $\chi (\phi) \equiv 1$, $\Lambda=0$, and $G_0(a)\equiv 0$ in Theorem \ref{Proposition I}.

Corollary \ref{PropositionIb} is a particular case of Corollary \ref{PropositionI} for a flat FLRW universe. 
\item[(C)]
\begin{cor}[Corollary of Proposition 2 of \cite{Miritzis:2003ym}]\label{tm}  Assuming that hypotheses i) and ii) of Theorem \ref{Proposition I} are satisfied, and $\chi (\phi) \equiv 1$, $\Lambda=0$, $G_0(a)\equiv 0$, $\rho_m=0$, then $\displaystyle{\lim_{t\rightarrow  \infty} y =0}$. 
\end{cor}
\textbf{Proof}.  Set $\chi (\phi) \equiv 1$ (minimal coupling), $\Lambda=0$, and $G_0(a)\equiv 0$ (flat FLRW universe), $\rho_m=0$ (vacuum)  in Theorem \ref{Proposition I}.
\end{enumerate}
\begin{thm}[Leon \& Franz-Silva, 2020]
\label{Theorem5.2} Assuming the following holds:  
\begin{enumerate}
\item $V(\phi)\in C^2(\mathbb{R}), V(\phi)\geq 0$, and $V(\phi)=0$ if and only if $\phi=0$. 
\item $V^{\prime}(\phi)$ is bounded on $A\subset\mathbb{R}$ if $V(\phi)$ is bounded on  $A$.
\item $\Lambda\geq 0$, and $G_0(a)\geq 0$ has a negative power-law functional form $G_0(a)=\frac{K^2}{a^p}, p>0$. 
\item $\chi(\phi)\in C^2(\mathbb{R})$, and for all $A\subset \mathbb{R}$ there is a non-negative constant $K_1$, possibly depending on $A$, such as $\left| \frac{\chi '(\phi)}{\chi (\phi)}\right|\leq K_1$ for all $\phi \in A$. 
\item $V^{\prime}(\phi)<0$ for $\phi<0$ and $V^{\prime}(\phi)>0$ for $\phi>0$.
\end{enumerate}
Then $\lim_{t\rightarrow\infty}\phi \in \{-\infty, 0, +\infty\}$. 
\end{thm}
\textbf{Proof}. As before, let $O^+(x_0)$ be the positive orbit that passes  through the regular point $x_0\in
\left\{(H, \rho_m, a, y, \phi)\in \Omega: H>0\right\}$ at the time $t_0$.
Using the same argument as in the proof of Theorem 
\ref{Proposition I}, $\lim_{t\rightarrow\infty}
H(t)=\eta$ through the orbit $O^+(x_0)$. Under the  hypotheses (i), (ii), (iii) and (iv) (see Theorem \ref{Proposition I}), it follows that $\lim_{t\rightarrow \infty} \left( \rho_m, y, \frac{K^2}{a^p}\right)=(0,0,0)$.

Supposing that  $3\eta^2=\Lambda,$ using the restriction  \eqref{Fried2bB} and Theorem \ref{Proposition I},  then, $\lim_{t\rightarrow\infty} V(\phi(t))=0$. As $V$ is continuous $V(\phi)=0\Leftrightarrow \phi=0$ this implies that 
$\lim_{t\rightarrow\infty} \phi(t)=0$.

Supposing that $3\eta^2>\Lambda$, using the restriction  \eqref{Fried2bB} and Theorem \ref{Proposition I}, it follows,
$\lim_{t\rightarrow\infty} V(\phi(t))=3\eta^2-\Lambda> 0$. Then, there is 
$t'$ such as $V(\phi)>(3\eta^2 -\Lambda)/2$ for all $t>t'$. From this fact, it follows that  $\phi$ cannot be zero for $t>t'$ due to  $\phi=0
\Leftrightarrow V(\phi)=0$. Thus, the sign of $\phi$ is invariant for all $t>t'$.

Assuming that $\phi$ is positive for all $t>t'$, and from the fact that $V$ is an increasing function of $\phi$ in $(0,+\infty)$, it follows that
$\lim_{t\rightarrow\infty} V(\phi(t))=(3\eta^2 -\Lambda)\leq
\lim_{\phi\rightarrow\infty} V(\phi)$.  By continuity and the monotonicity of $V$, it follows that this equality holds if and only if $\lim_{t\rightarrow\infty} \phi(t)=+\infty$.

If $\lim_{t\rightarrow\infty} V(\phi(t))<
\lim_{\phi\rightarrow\infty} V(\phi)$, then, there exists
$\bar{\phi}> 0$ such as: $$ \lim_{t\rightarrow\infty}
V(\phi(t))=V(\bar{\phi}).$$   Due to the fact that $V$ is strictly increasing and continuous, then $ \lim_{t\rightarrow\infty} \phi=\bar{\phi}.$ 

 Taking the limit $t\rightarrow\infty$ on Equation \eqref{13EQ} and using that $V(\phi)\in C^2(\mathbb{R})$, it follows that:
$$ \lim_{t\rightarrow\infty}\frac{d y}{d
t}=-V'(\bar{\phi})<0.$$   Hence, there exists $t''> t'$ such that
$\frac{d y}{d t}<-V'(\bar{\phi})/2$  for all $t\geq t'' $. This implies
$$ y(t)- y(t'')=\int_{t''}^{t}\left(\frac{d y}{d t}\right)
dt \leq -\frac{ V'(\bar{\phi})}{2}(t- t'').$$  That is, $y(t)$ takes negative values large enough as $t\rightarrow \infty$, which is impossible because $\lim_{t\rightarrow\infty}y(t)=0$. Henceforth, if $\phi>0$ for all $t>t'$, then
$\lim_{t\rightarrow\infty}\phi=+\infty$. In the same way, for $\phi<0$ and
for all $t>t'$, then $\lim_{t\rightarrow\infty}\phi=-\infty$. $\blacksquare$

If initially
$3H(t_0)^2<\min\left\{\lim_{\phi\rightarrow\infty}V(\phi),
\lim_{\phi\rightarrow -\infty}V(\phi)\right\}$,  then
$\lim_{t\rightarrow\infty}H(t)=\sqrt{\frac{\Lambda}{3}}$. Indeed, the conclusion of Theorem \ref{Theorem5.2} is 
$\lim_{t\rightarrow\infty}\phi \in \{-\infty, 0, +\infty\}$. If $\lim_{t\rightarrow\infty}\phi=+\infty$, from the restriction \eqref{Rachd2}, it follows that:
$$ 3 \eta^2-\Lambda=\lim_{t\rightarrow\infty}V(\phi(t))=
\lim_{\phi\rightarrow\infty} V(\phi)>3 H(t_0)^2,$$  in contradiction with the fact that $H(t)$ is decreasing and $H(t_0)\geq
\eta, \Lambda\geq 0$. In the same way, the assumption $\lim_{t\rightarrow\infty}\phi=-\infty$
leads to a contradiction. Thus,  $\lim_{t\rightarrow\infty}\phi=0$
which implies $\lim_{t\rightarrow\infty}V(\phi(t))=0$, and from \eqref{Rachd2} it follows that $\lim_{t\rightarrow\infty}H(t)=\sqrt{\frac{\Lambda}{3}}$. 

	Now, it will be shown how Theorem \ref{Theorem5.2} generalizes previous results in the following corollaries:
\begin{enumerate}
\item[(A)]
\begin{cor}[Leon \& Franz-Silva, 2020]\label{Proposition II}
Under the hypotheses (i), (ii), (iii) and (v) of Theorem \ref{Theorem5.2}, and setting $\chi(\phi)\equiv 1$, then  $\lim_{t\rightarrow\infty}\phi \in \{-\infty, 0, +\infty\}$. 
\end{cor}
\textbf{Proof}. Set $\chi (\phi) \equiv 1$ (minimal coupling) in Theorem \ref{Theorem5.2}.

\item[(B)]
\begin{cor}[Miritzis 2003. Proposition 3, \cite{Miritzis:2003ym}] 
\label{PropositionII}
Under the hypotheses (i), (ii) and (v) of Theorem \ref{Theorem5.2}, and setting $\chi(\phi)\equiv 1$, $\Lambda=0$, and $G_0(a)\equiv 0$, 
then $\displaystyle{\lim_{t\rightarrow  \infty} \phi(t) \in \lbrace -\infty , 0 , + \infty \rbrace }$. 
 \end{cor}
 \textbf{Proof}. Set $\chi (\phi) \equiv 1$, $\Lambda=0$, and $G_0(a)\equiv 0$ in Theorem \ref{Theorem5.2}.
\item[(C)]
\begin{cor}[Corollary of Proposition 3 of \cite{Miritzis:2003ym}]\label{thm2.1} Under the hypotheses (i), (ii) and (v) of Theorem \ref{Theorem5.2}, and setting $\chi (\phi) \equiv 1$, $G_0(a)\equiv 0$, $\Lambda=0$, and $\rho_m=0$, then $\displaystyle{\lim_{t\rightarrow  \infty} \phi(t) \in \lbrace -\infty , 0 , + \infty \rbrace }$. 
 \end{cor}
 \textbf{Proof}. Set  $\chi (\phi) \equiv 1$ (minimal coupling),  $\Lambda=0$, $G_0(a)\equiv 0$ (flat FLRW universe) and $\rho_m=0$ (vacuum)  in the Theorem \ref{Theorem5.2}.
\end{enumerate}
\begin{thm}[Leon \& Franz-Silva, 2020]
\label{Theorem5.3}
Assuming the following holds: 
\begin{enumerate}
    \item $V(\phi)\in C^2(\mathbb{R})$, $V(\phi)\geq 0$ and $\lim_{\phi \rightarrow -\infty}V(\phi)=+\infty$.
    \item $V^{\prime}(\phi)$ is continuous and $V^{\prime}(\phi)<0$.
    \item $V^{\prime}(\phi)$ is bounded on $A\subset\mathbb{R}$ if $V(\phi)$ is bounded on  $A$.
    \item $\Lambda \geq 0$ and $G_0(a)=\frac{K^2}{a^p}, p>0$.
    \item $\chi(\phi)\in C^2(\mathbb{R})$ such as for all $A\subset \mathbb{R}$ there is a constant $K_1$, possibly depending on $A$, such as $\left| \frac{\chi '(\phi)}{\chi (\phi)}\right|<K_1$ for all $\phi \in A$.
\end{enumerate}
Then,  $\lim_{t\rightarrow \infty} \left( \rho_m, y, \frac{K^2}{a^p}\right)=(0,0,0)$ and
$\lim_{t\rightarrow\infty}\phi =+\infty$. 
\end{thm}
\textbf{Proof}.
As before, let $O^+(x_0)$ be the positive orbit that passes  through the regular point $x_0\in
\left\{(H, \rho_m, a, y, \phi)\in \Omega: H>0\right\}$ at the time $t_0$. From equation  \eqref{consb}, it follows that the set  $\rho_m>0$ is invariant for the flow of \eqref{consb} with the restriction \eqref{Fried2bB} through the orbit $O^+(x_0)$. That is, $\rho_m$ is different from zero if initially $\rho_m(t_0)$ is so. This implies  $H$ is never zero because of equation \eqref{Fried2bB}, i.e., $3 H(t)^2\geq \rho_m(t)>0$ for all $t>t_0$.  Then,
$H$ is always non-negative if initially it is non-negative. Furthermore, from equation \eqref{Rachd2}, it follows that  $H$ is decreasing and non-negative, then
$\lim_{t\rightarrow\infty} H(t)=\eta\geq 0$ and 
$$ \int_{t_0}^\infty \left(\frac{1}{6} K^2 p a(s)^{-p}+\frac{1}{2}
   \gamma  \rho_m (s)+\frac{1}{2} y(s)^2\right) ds = H(t_0)-\eta<+\infty.$$   As in Theorem \ref{Proposition
I}, the total derivative of $f(t)=\left(\frac{1}{6} K^2 p a(t)^{-p}+\frac{1}{2}
   \gamma  \rho_m (t)+\frac{1}{2} y(t)^2\right)$ is bounded, and the improper integral $\int_{t_0}^\infty f(t) dt$ is convergent. Then, $\lim_{t\rightarrow +\infty} f(t)=0$, together with the non-negativeness of each term of $f(t)$ implies: $$ \lim_{t\rightarrow \infty} \left(\rho_m, y, \frac{K^2}{a^p}\right)=(0,0,0).$$  
It can be proved that $\lim_{t\rightarrow\infty}\phi=+\infty$ in the same way as it was proved in Theorem \ref{Proposition II}. From equation \eqref{Fried2bB} it follows that $\lim_{t\rightarrow\infty} V(\phi)=3\eta^2-\Lambda$.  The function $V$ is strictly decreasing with respect to $\phi$, then $V(\phi)>\lim_{\phi\rightarrow\infty} V(\phi)$ for all $\phi$. Hence,  $\lim_{t\rightarrow\infty} V(\phi(t))\geq
\lim_{\phi\rightarrow\infty} V(\phi)$. Thus, there are two cases to be considered: 
 \begin{enumerate}
   \item If $\lim_{t\rightarrow\infty} V(\phi(t))= \lim_{\phi\rightarrow\infty} V(\phi)$, then,
 $\lim_{t\rightarrow\infty}\phi=+\infty$ by continuity of $V$.
   \item If  $ \lim_{t\rightarrow\infty} V(\phi(t))> \lim_{\phi\rightarrow\infty} V(\phi)$, then,  there is a unique $\bar{\phi}$ such as:
   $$ \lim_{t\rightarrow\infty} V(\phi(t))=V(\bar{\phi}),$$ by continuity and monotonicity of $V$. By continuity, it follows that $ \lim_{t\rightarrow\infty} \phi=\bar{\phi}$.  
\newline Taking limit as $\phi \rightarrow \infty$ in Equation \eqref{13EQ} and using the hypothesis $V(\phi)\in C^2(\mathbb{R})$, it follows that
$$ \lim_{t\rightarrow\infty}\frac{d y}{d 
t}=-V'(\bar{\phi})>0.$$   Hence,  there exists $t'$ such that
$\frac{d y}{d t}>-V'(\bar{\phi})/2$ for all $ t\geq t'$. Therefore, 
$$ y(t)- y(t')
>-\frac{V'(\bar{\phi})}{2}(t- t'),$$  which is impossible because
$\lim_{t\rightarrow\infty}y(t)=0$. Finally,
$\lim_{t\rightarrow\infty}\phi=+\infty$. $\blacksquare$
 \end{enumerate} 
 Additionally, if 
$\lim_{\phi\rightarrow\infty} V(\phi)=0$, then
$H\rightarrow \sqrt{\frac{\Lambda}{3}}$ as $t \rightarrow\infty$.

	Now, it will be shown how Theorem \ref{Theorem5.3} generalizes previous results in the following corollaries: 
\begin{enumerate}
    \item[(A)] 
    \begin{cor}[Leon \& Franz-Silva, 2020]
\label{Proposition III}
Under the hypotheses (i), (ii), (iii) and (iv) of Theorem \ref{Theorem5.3}, and assuming  $\chi (\phi) \equiv 1$, then, 
$ \lim_{t\rightarrow \infty} \left( \rho_m, y, \frac{K^2}{a^p}\right)=(0,0,0)$ and $\lim_{t\rightarrow\infty}(a, \phi) =(+\infty,+\infty)$. 
\end{cor}
\textbf{Proof}. Set  $\chi (\phi) \equiv 1$ in Theorem \ref{Theorem5.3}.

\item[(B)]  
\begin{cor}[Miritzis 2003. Proposition 4, \cite{Miritzis:2003ym}]
\label{Prop4Miritzis} Under the hypotheses (i), (ii) and (iii) of Theorem \ref{Theorem5.3}, and setting $\chi (\phi) \equiv 1$, $\Lambda=0$ and $G_0(a)\equiv 0$, then,  $\lim_{t \rightarrow  + \infty} y(t)=\lim_{t \rightarrow  + \infty} \rho_m(t) =0$ and $\lim_{t \rightarrow  + \infty} \phi (t)= +\infty$.
 \end{cor}
 \textbf{Proof}. Set  $\chi (\phi) \equiv 1$ (minimal coupling), $\Lambda=0$, and choose $G_0(a)\equiv 0$ (flat FLRW universe) and $\Lambda=0$ in Theorem \ref{Theorem5.3}.
 
\item[(C)] 
 \begin{cor}[Corollary of Proposition 4 of \cite{Miritzis:2003ym}]\label{thm2.2} Under the hypotheses (i) and (ii) of Theorem \ref{Theorem5.3}, and setting  $\chi (\phi) \equiv 1$, $\Lambda=0$,  $G_0(a)\equiv 0$ and $\rho_m=0$, then $\lim_{t \rightarrow  + \infty} y(t) =0$ and $\lim_{t \rightarrow  + \infty} \phi (t)= +\infty$.
 \end{cor}
 \textbf{Proof}.  Set $\chi (\phi) \equiv 1$ (minimal coupling), $\Lambda=0$, $G_0(a)\equiv 0$ (flat FLRW universe) and $\rho_m=0$ (vacuum)  in Theorem \ref{Theorem5.3}.
\end{enumerate}
Finally, the next theorem is presented: 
\begin{thm}[Leon \& Franz-Silva, 2020]\label{thmIIIFINAL} Assuming the following holds 
\begin{enumerate}
\item[(i)] $V(\phi)\in C^2(\mathbb{R})$ such that the possibly empty set $\{\phi: V(\phi)<0\}$ is bounded 
\item[(ii)] and the possibly empty set of singular points of  $V(\phi)$ is finite.
\item[(iii)] $\phi_*$ is a strict minimum possibly degenerated of $V(\phi)$ with a positive critical value.
\item[(iv)] $\Lambda= 0$, and $G_0(a)\geq 0$ such as $\frac{a  G_{0}'(a)}{G_{0}(a)}\leq -p<0$, for all $a>0$. 
\end{enumerate}
 Then ${\bf
p}_*:=(\phi, y, \rho_m, H)=\left(\phi_*, 0, 0, \sqrt{\frac{V(\phi_*)}{3}}\right)$ is an asymptotically stable equilibrium point.
\end{thm}
\textbf{Proof}. Defining 
\begin{equation}
\label{defW}
W(\phi, y, \rho_m, H)=H^2-\frac{1}{3}\left(\frac{1}{2}y^2+V(\phi)+\rho_m\right):= \frac{1}{3}G_0(a),
\end{equation}
which satisfies \footnote{Observing that in \cite{Giambo:2009byn} where the case $G_0(a)=-\frac{3k}{a^2}$ was studied, it leads to $\dot W=-2 H W$.}
\begin{equation}
\label{16Eq}
\dot W=H W \frac{a  G_{0}'(a)}{G_{0}(a)} <-p H W.
\end{equation}
Therefore, $W$ is a non-negative and decreasing function of $t$ for $H> 0$. 

Defining: 
\begin{equation}
\label{17Eq}
\epsilon =\frac{1}{2}y^2 + V(\phi) + \rho_m, \quad \dot\epsilon =-3 H \left(\gamma  \rho_m +y^2\right).
\end{equation}
These imply that $\epsilon$ is decreasing for $H>0$ and $\rho_m\geq 0$. 

Firstly, it is assumed that $V(\phi_*)>0$.  Letting 
$\tilde{V}>V(\phi_*)$ be a regular value of $V$ such that the connected component of $V^{-1}\left((-\infty,\tilde{V}]\right)$
that contains $\phi_*$ is a compact set in $\mathbb{R}$. Denoting this set by $A$  and defining  $\Psi$ as:
$$ \Psi=\left\{(\phi,y, \rho_m,H)\in\mathbb{R}^4: \phi\in A,
\epsilon \leq \tilde{V}, \rho_m\geq 0, W(\phi, y, \rho_m, H)\in[0, \bar{W}]\right\},$$  where $\bar{W}$ is positive. It can be proved that $\Psi$ is a compact set as follows:
\begin{enumerate}
\item $\Psi$ is a closed set in $\mathbb{R}^4$.
\item $V(\phi_*)\leq V(\phi) \leq \tilde{V}$, for all $\phi\in A$.
\item Since $\frac{1}{2} y^2 + V(\phi_*)\leq \frac{1}{2} y^2 + V(\phi) + \rho_m= \epsilon \leq \tilde{V}$. Therefore, it  follows that $y$ is bounded.
\item From $\rho_m\leq \tilde{V} - \frac{1}{2} y^2 - V(\phi) \leq \tilde{V} -  V(\phi_*)$, it is a consequence that $\rho_m$ is bounded. 
\item From \eqref{defW}, and due to the above facts, it also follows that:
\begin{equation}
\frac{V(\phi_*)}{3}\leq\frac{V(\phi)}{3}\leq H^2=W+\frac{1}{3}\left(\frac{1}{2}y^2+V(\phi)+\rho_m\right) \leq \bar{W} + \frac{\tilde{V}}{3}.
\end{equation} That is, $H$ is also bounded. 
\end{enumerate}
Defining $\Psi_+\subseteq \Psi$, the connected component of  $\Psi$
containing ${\bf p}_*$.  Then, following the same arguments as in 
\cite{Giambo:2009byn,Giambo:2008ck} it is proved that $\Psi_+$ is positively invariant with respect to \eqref{Non_min2}. Letting  $\mathbf{x}(t)$ to be any solution starting at $\Psi_+$, and defining $\bar{t}=\sup\left\{t>0:
H(t)>0\right\}\in\mathbb{R}\cup\{+\infty\}$, it follows that $t<\bar{t}$. Equations \eqref{16Eq} and \eqref{17Eq} imply that both $W$ and $\epsilon$ decrease.  
Moreover, it is assumed that there exists $t<\bar{t}$ such as $\phi(t)\notin A$. Hence, $V(\phi(t))>\tilde{V}$.  However 
\begin{equation*}
\tilde{V}< V(\phi(t)) \leq \frac{1}{2}y(t)^2 + V(\phi(t)) + \rho_m(t)= \epsilon(t) \leq \tilde{V},
\end{equation*}
which is a contradiction. Therefore,  $\phi(t)\in A, \quad \forall t<\bar{t}$. But, $W\geq 0$ along with the flow under \eqref{Non_min2}, due to  $G_0(a)\geq 0$, and so by hypothesis it follows that: 
\begin{equation*}
H(t)^2 \geq \frac{1}{3}\left(\frac{1}{2}y(t)^2+V(\phi(t))+\rho_m(t)\right)\geq  \underbrace{\frac{V(\phi(t))}{3}\geq \frac{V(\phi_*)}{3}}_{\text{Because}\; \phi(t)\in A, \quad \forall t<\bar{t}}.
\end{equation*}
 That is, as long as $H$ remains positive, it is strictly bounded away from zero and thus $\bar{t}=+\infty$. Therefore, $\Psi_+$ satisfies the hypothesis of LaSalle's invariance Theorem \cite{LaSalle,wiggins}. Considering the monotonic functions  $\epsilon$ and $W$ defined on $\Psi_+$, it follows that any solution with initial state on $\Psi_+$ must be such that $H y^2 \rightarrow 0, H \gamma \rho_m \rightarrow 0$ as $t\rightarrow +\infty$.  Since $H$ is strictly bounded away from zero on  $\Psi_+$ and $\gamma>0$, it follows that 
$y\rightarrow 0$, $\rho_m\rightarrow 0$  and 
$H^2-\frac{V(\phi)}{3}\rightarrow 0$ as $t\rightarrow +\infty$.   
From 
hypotheses $G_{0}(a)\geq 0$ and  $\frac{a  G_{0}'(a)}{G_{0}(a)}\leq -p<0$, for all $a>0$, it follows that:
\begin{equation*}
\dot{H}=-\frac{1}{2}\left(\gamma \rho_m+y^2\right)+\frac{1}{6}a G_0'(a) \leq -\frac{1}{2}\left(\gamma \rho_m+y^2\right)-\frac{p}{6} G_0(a) \leq 0. 
\end{equation*} 
Given that $H$ is monotonically decreasing, positive and bounded, it must have a limit. This implies that 
$V(\phi)$ must also have a limit, and this must be $V(\phi_*)$.  Otherwise,  $V^{\prime}(\phi)$ would tend to a non zero value, and as well as the right hand side of \eqref{13EQ} which is a contradiction. Therefore, the solution tends to ${\bf p}_*$.

If $V(\phi_*)=0$, the set $\Psi$ is connected and  $\Psi_{+}$ is chosen as the subset of $\Psi$ with $H\geq0$. The unique equilibrium point on  $\Psi_{+}$ with  $H=0$ is then the equilibrium point
${\bf p}_*$, then, if $H(t)\rightarrow 0$, the solution is forced to tend to the equilibrium point due to $H$ is monotonic.  On the contrary, if $H(t)$ tends to a positive number, as before, it verifies that  $y\rightarrow 0, \rho_m\rightarrow 0, W \rightarrow 0,\,V(\phi)\rightarrow V(\phi_*)=0$, and hence $H$ will necessarily tend to zero. $\blacksquare$

\subsubsection{Case of positive curvature $k=1$.} 
\label{positive-k}
For $k=+1, W<0$, it cannot be guaranteed the monotony of $H$, so the previous arguments are adapted  in exactly the same way as in \cite{Giambo:2008ck}.  That is, $\phi_*$ is a local minimum of $V(\phi)$ with $V(\phi_*)>0$.  The value 
$\tilde{V}>V(\phi_*)$ is a regular value of $V$ so the connected  component of $V^{-1}\left((-\infty,\tilde{V}]\right)$
 contains $\phi_*$ as the only critical point of $V$, and is a compact set in $\mathbb{R}$. It is considered a solution $\mathbf{x}(t)=(\phi(t), y(t), \rho_m(t), H(t))$ such that $\frac{1}{2} y(0)^2 +V(\phi(0)) + \rho_m(0)\leq \tilde{V}$ and let $\bar{W}<0$, a value to determine, to act as a lower bound for $W$. 
Taking the initial condition near the equilibrium point ${\bf
p}_*$, then $H(0)>0$; since $W(0)>\bar{W}$, from the equations $\dot W= H W \frac{a  G_{0}'(a)}{G_{0}(a)}\geq 0$ and $\dot \epsilon =-3 H \left(\gamma  \rho_m +y^2\right)\leq 0$, it follows that $W\geq \bar{W}$ ($W$ now, it will be monotonic increasing and negative) and $\epsilon\leq \tilde{V}$. The last inequality implies that $V(\phi(t))\leq \tilde{V}$. This implies that $\phi(t)$ satisfies $V(\phi_*)\leq V(\phi(t))\leq \tilde{V}$. Then,
\begin{equation*}
H^2= \frac{\epsilon}{3}  + \frac{W}{3} \geq \frac{V(\phi_*)}{3}+ \frac{\bar{W}}{3}= H_*^2 +\frac{\bar{W}}{3} \implies
H\geq  \bar{H}:= \left(\sqrt{1+\frac{\bar{W}}{3 {H_*}^2}}\right) H_*, 
\end{equation*}
where $H_*=\sqrt{\frac{V(\phi_*)}{3}}$.
Choosing $\bar{W}$ small enough such as $H_*^2>-\frac{\bar{W}}{3}$, it follows that $H(0)>0 \implies H(t)\geq \bar{H} >0$.
That is, $H$ is bounded away from zero, which is combined with the monotony of $W$ and  $\epsilon$, and using the LaSalle's invariance Theorem as in the case $W\geq 0$ leads to $y\rightarrow 0, \rho_m\rightarrow 0, W \rightarrow 0$ as $t\rightarrow \infty$, and the equilibrium point ${\bf
p}_*$ is approached asymptotically.  

If $V(\phi_*)=0$, the equilibrium point satisfies $H_*=0$, and the nearby solutions may re-collapse as $H$ changes sign.  Collapsing models were exhaustively studied, e.g., in \cite{Giambo:2009zza,Giambo:2015tja,Giambo:2014jfa,Giambo:2013bya,Giambo:2009zz,Giambo:2008sa,Giambo:2008ya,Giambo:2005se,Giambo:2002tp,Giambo:2001wi} for a wide class of self--interacting,  self-gravitating homogeneous scalar field models.

In the next sections  some examples that satisfy the hypotheses of the Theorems that were proved in Section \ref{SECT:Non_min} are presented, as well as some examples that do not satisfy one or more hypotheses of these Theorems to obtain some counterexamples.  
 
\subsection{Generalized harmonic potential $
V(\phi)= \mu^3 \left[\frac{\phi^2}{\mu} + b f \cos\left(\delta + \frac{\phi}{f}\right)\right]$, $b\neq 0$ in vacuum.}
\label{Sect2.4}

\begin{figure}[t]
\begin{center}
	\subfigure[$(b, f, \delta, \mu)= (0.1, 0.33, 0, 0.9)$]{\includegraphics[scale=0.5]{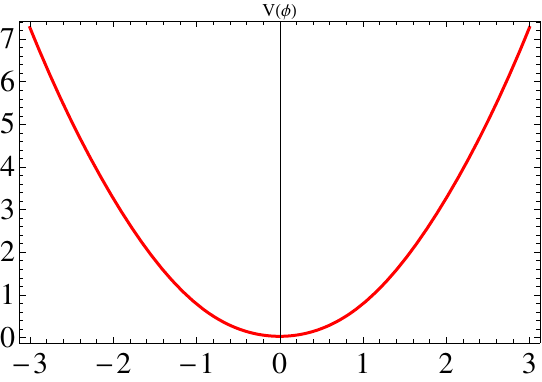} } \hspace{2cm}
	\subfigure[$(b, f, \delta, \mu)= (0.1, 0.33, 0, 0.9)$]{\includegraphics[scale=0.5]{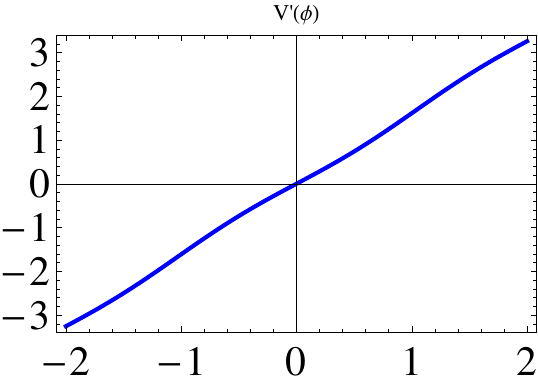} }
	\subfigure[$(b, 
   f, \delta, \mu)= (0.99, 0.09, 0, 0.9)$]{\includegraphics[scale=0.5]{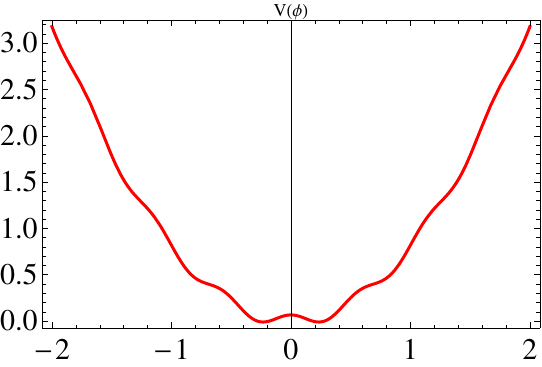}} \hspace{2cm}
	\subfigure[$(b, 
   f, \delta, \mu)= (0.99, 0.09, 0, 0.9)$]{\includegraphics[scale=0.5]{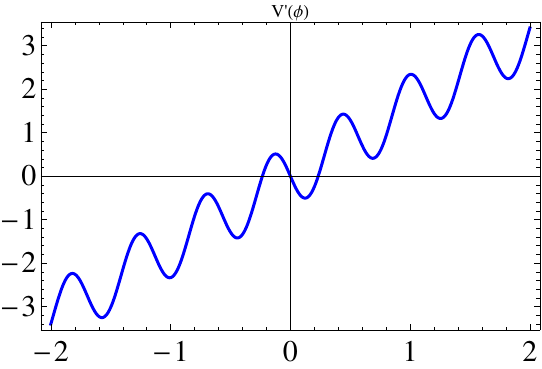} }
\end{center}\caption{\label{PlotMonodromy-Potential} Generalized harmonic potential $
V(\phi)= \mu^3 \left[\frac{\phi^2}{\mu} + b f \cos\left(\delta + \frac{\phi}{f}\right)\right]$ and its derivative.} 
\end{figure}

The focus of this section is the qualitative analysis  of a scalar-field cosmology with a generalized harmonic potential 
\begin{equation}
\label{harmonic1}
  V(\phi)= \mu^3 \left[\frac{\phi^2}{\mu} + b f \cos\left(\delta + \frac{\phi}{f}\right)\right], \quad b\neq 0, 
\end{equation} in a vacuum.
In figure \ref{PlotMonodromy-Potential} the potential $V(\phi)$ and its derivative $V^{\prime}(\phi)$ are represented for some values of the parameters $(b, f, \delta, \mu)$. The condition for the existence of a local minimum at the origin is $\delta=0, \mu ^3 \left(\frac{2}{\mu }-\frac{b}{f}\right)>0$; with $V(0)=b f \mu^3$. 

The condition for the existence of a local maximum at the origin is $\delta=0, \mu ^3 \left(\frac{2}{\mu }-\frac{b}{f}\right)<0$; with $V(0)=b f \mu^3$. For $\delta= 0,\mu= \frac{2 f}{b}$, $\phi= 0$ is a degenerated local minimum of order two with $V(0)=\frac{8 f^4}{b^2}$. 

For simplicity,  the re-scaling  $(y \equiv \dot \phi, \phi) \rightarrow (u, v)$ given by:
\begin{equation}
\label{EQ:19}
    u=\frac{\dot\phi}{\sqrt{2 \rho_c}}, \quad v= \frac{\phi}{f}, 
\end{equation}
is implemented where the parameter $\rho_c>0$ will be chosen conveniently.

Using this parametrization, the Friedmann equation can be rewritten as 
\begin{equation}
  \frac{3 H^2}{\rho_c}=\left[u^2+ \frac{\mu^3 f}{\rho_c}\left(\frac{f v^2}{\mu} + b \cos\left(\delta + v\right)\right)\right].
    \end{equation}
To describe an expanding universe the positive solution for $H$ from the previous equation is chosen. Hence, by introducing $\tau =\frac{\sqrt{2 \rho_c}}{f} t$, and
redefining constants $\rho_{c}= \frac{1}{2} b f \mu ^3>0, \quad k=\frac{2 f}{b \mu}>0$,
 the following equations are obtained:   
\begin{align}
&\frac{d u}{d \tau}=-\frac{\sqrt{6}}{4} b k \mu  u \sqrt{k v^2+u^2+2 \cos (\delta +v)}-k \mu  v+\sin (\delta +v)), \quad\frac{d v}{d\tau}=u.  \label{0monodronomyeqs1ab}
\end{align}
\newline
The origin $(u,v)=(0,0)$ is an equilibrium point if $\delta=0$. The eigenvalues of the linearization matrix of \eqref{0monodronomyeqs1ab} are \\ 
$\left\{\frac{1}{4} \left(-\sqrt{k \mu  \left(3 b^2 k \mu -16\right)+16}-\sqrt{3} b k \mu \right),\frac{1}{4} \left(\sqrt{k \mu  \left(3
   b^2 k \mu -16\right)+16}-\sqrt{3} b k \mu \right)\right\}$. 

The origin is a sink for:
   $\mu >0, k>\frac{1}{\mu }, b\geq \frac{4}{3} \sqrt{\frac{3 k \mu -3}{k^2 \mu ^2}}$. When $\delta=0$ and $\mu =0$, or $k \mu <1$ the origin is a saddle. 
\newline
Now, for $k\neq 0$ and $|k \mu v_c|\leq 1$, the following are equilibrium points of \eqref{0monodronomyeqs1ab}, $(u,v)=(0,v_c)$ where  $v_c$ are the roots of the transcendental equation 
$\sin (\delta +v)-k \mu  v=0$.  To obtain a real valued linearization matrix, it is additionally required that
$\frac{3 v_c \sin (\delta +v_c)}{2 \mu }+3 \cos (\delta +v_c)\geq 0$. 
\newline
If $\delta=0$ and $|k \mu v_c|>1$, there are no equilibrium points other than the origin.  
\newline
In general, for $\delta \neq 0$ the system \eqref{0monodronomyeqs1ab} admits no equilibrium points $(u,v)=(0,v_c)$, other than the origin for $|k \mu v_c|>1$. 
\newline
If $|k \mu v_c|\leq 1$, the following are equilibrium points of \eqref{0monodronomyeqs1ab}, $(u,v)=(0,v_c)$ where $v_c$ are the roots of the transcendental equation $\sin (\delta +v)-k \mu  v=0$.  
\newline
For $\delta\neq 0$ and $|k \mu v_c|\leq 1$, it is deduced that
    $\delta = -v_c+ \arcsin\left(k \mu v_c\right)$, and  the eigenvalues \\
$\Big\{-\frac{\sqrt{k \mu  \left(3 b^2 k \mu  \left(2 \sqrt{1-k^2 \mu ^2 v_c^2}+k v_c^2\right)-32\right)+32 \sqrt{1-k^2 \mu ^2 v_c^2}}+b k \mu 
   \sqrt{6 \sqrt{1-k^2 \mu ^2 v_c^2}+3 k v_c^2}}{4 \sqrt{2}}$,\\
	$\frac{\sqrt{k \mu  \left(3 b^2 k \mu  \left(2 \sqrt{1-k^2 \mu ^2 v_c^2}+k
   v_c^2\right)-32\right)+32 \sqrt{1-k^2 \mu ^2 v_c^2}}-b k \mu  \sqrt{6 \sqrt{1-k^2 \mu ^2 v_c^2}+3 k v_c^2}}{4 \sqrt{2}}\Big\}$ are obtained. 

\begin{figure}[t]
\begin{center}
\subfigure[\label{monodromy-potential-1-b} $(b, f, \delta, \mu)= (0.1, 0.33, 0, 0.9)$]{\includegraphics[scale=0.4]{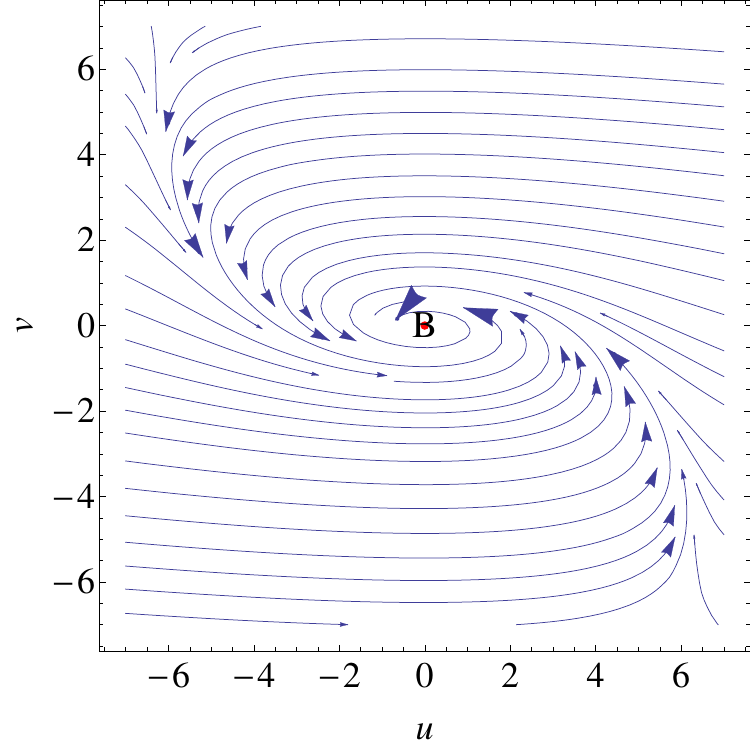}}\hspace{2cm}
\subfigure[\label{monodromy-potential-1-a}  $(b, f, \delta, \mu)= (0.99, 0.09, 0, 0.9)$]{\includegraphics[scale=0.4]{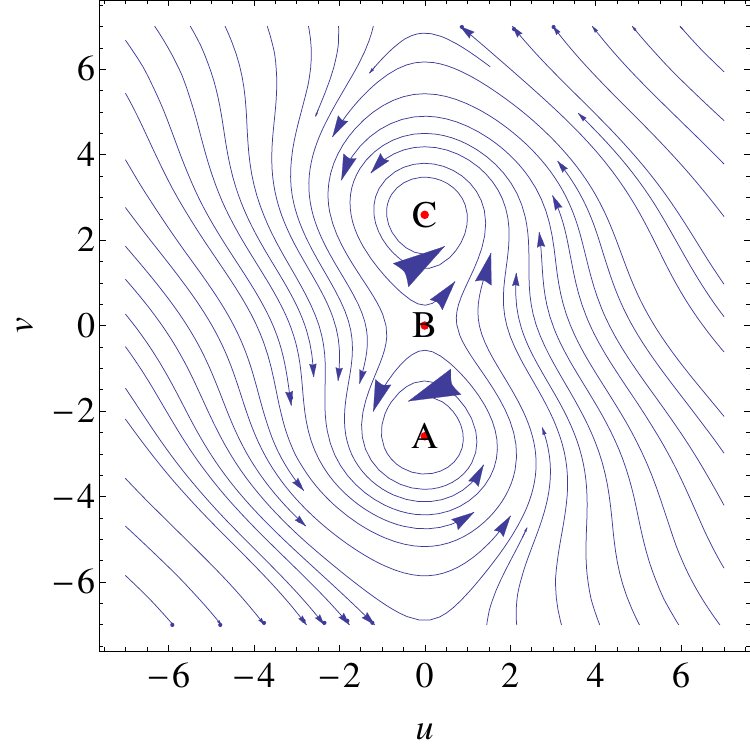}}
\caption{Phase portrait of equations    \eqref{0monodronomyeqs1ab} for some values of parameters $(b, f, \delta, \mu)$.}
\end{center}
\end{figure}

For the choice of parameters $(b, f, \delta, \mu)= (0.1, 0.33, 0, 0.9)$, it follows that
   $\rho_c=\frac{24057}{2000000}\approx 0.0120285, k=\frac{22}{3}\approx 7.33333$. 
The only equilibrium point is the origin with eigenvalues $\{-0.285788+2.34911 i,-0.285788-2.34911 i\}$, which is an stable spiral. In figure \ref{monodromy-potential-1-b}, some orbits of the flow of \eqref{0monodronomyeqs1ab} are depicted for the choice of parameters $(b, 
   f, \delta, \mu)=(0.1, 0.33, 0, 0.9)$.
For this choice of parameters  the hypotheses and the results of Theorems \ref{tm} and \ref{thm2.1} are verified. That is $\lim_{t\rightarrow \infty } \dot\phi=0$ and $\lim_{t\rightarrow \infty } \phi=0$.

Setting the values $(b, 
   f, \delta, \mu)= (0.99, 0.09, 0, 0.9)$, it is obtained that
   $\rho_c=\frac{649539}{20000000}\approx 0.032477, k=\frac{20}{99}\approx 0.20202$. 
The transcendental equation is $\frac{2 v}{11}-\sin (v)=0$. 
Therefore, there are three equilibrium points:
\begin{enumerate}
    \item $A:=(u,v)=(0, -2.64078)$. The linearization matrix is complex-valued with eigenvalues $\{ 0.997194 i,  -1.06199 i\}$. 
    \item $B:=(u,v)=(0,0)$ with eigenvalues $\{-0.985828,0.829944\}$. It is a saddle. 
       \item $C:=(u,v)=(0, 2.64078)$. The linearization matrix is complex-valued  with eigenvalues $\{0.997194 i,  -1.06199 i\}$. 
\end{enumerate}
In this case, the potential has negative values at the stable equilibrium points. It is well known that a negative constant potential generates an equilibrium state which is just the Anti - de Sitter (AdS) equilibrium solution.

In figure \ref{monodromy-potential-1-a}, some orbits of the flow of \eqref{0monodronomyeqs1ab}  are depicted for  $(b, f, \delta, \mu)= (0.99, 0.09, 0, 0.9)$. For these choices of parameters the hypotheses 
{\it{$V(\phi)\geq 0$ and $V(\phi)=0$, if and only if $\phi=0$}} of Theorem \ref{tm} are not satisfied, but the result $\lim_{t\rightarrow +\infty} \dot \phi =0$ holds. The hypotheses {\it{$V(\phi)\geq 0$ and $V(\phi)=0$, if and only if $\phi=0$}} and {\it{$V^{\prime}(\phi)<0$ for $\phi<0$ and $V^{\prime}(\phi)>0$ for $\phi>0$}} of Theorem \ref{thm2.1} are not fulfilled, and $\lim_{t\rightarrow +\infty}\phi$ can be finite (rather than zero or infinity). Recalling that, this Theorem relies on the former hypothesis. Finally,  the case when the hypotheses {\it{$V(\phi)\geq 0$ and $V^{\prime}(\phi)<0\quad \forall \phi\in\mathbb{R}$}} of \ref{thm2.2} are not satisfied, and $\lim_{t\rightarrow +\infty}\dot\phi=0, \lim_{t\rightarrow +\infty}\phi<\infty$ is illustrated.

\subsection{Generalized harmonic potential $
V(\phi)= \mu ^3 \left[b f \left(\cos (\delta )-\cos \left(\delta +\frac{\phi }{f}\right)\right)+\frac{\phi ^2}{\mu}\right]
$, $b\neq 0$, in vacuum.}
\label{Sect.2.5}
In this section, the qualitative analysis of a scalar-field cosmology with generalized harmonic potential 
\begin{equation}
\label{harmonic2}
V(\phi)= \mu ^3 \left[b f \left(\cos (\delta )-\cos \left(\delta +\frac{\phi }{f}\right)\right)+\frac{\phi ^2}{\mu}\right], \quad b\neq 0,
\end{equation}
in a vacuum is presented.

\begin{figure}[t!]
\begin{center}
	\subfigure[$(b, f, \delta, \mu)= (0.1, 0.33, 0, 0.9)$]{\includegraphics[scale=0.5]{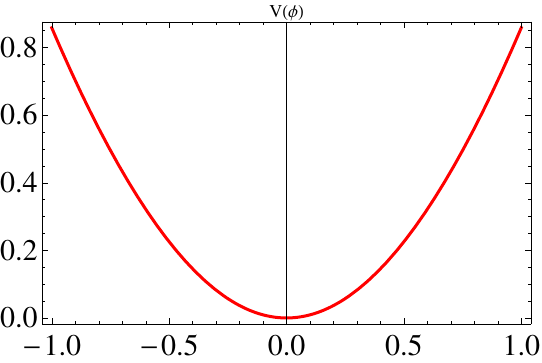} }\hspace{2cm}
	\subfigure[$(b, f, \delta, \mu)= (0.1, 0.33, 0, 0.9)$]{\includegraphics[scale=0.5]{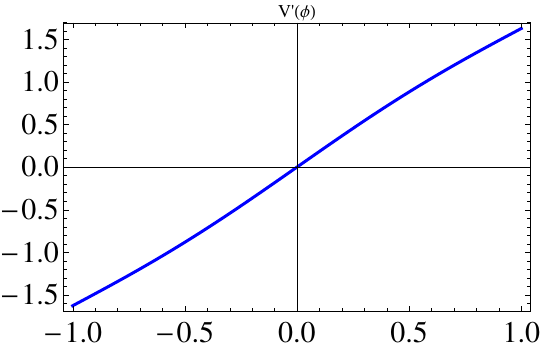} }
			\subfigure[$(b, f, \delta, \mu)= (0.99, 0.09, 0, 0.9)$]{\includegraphics[scale=0.5]{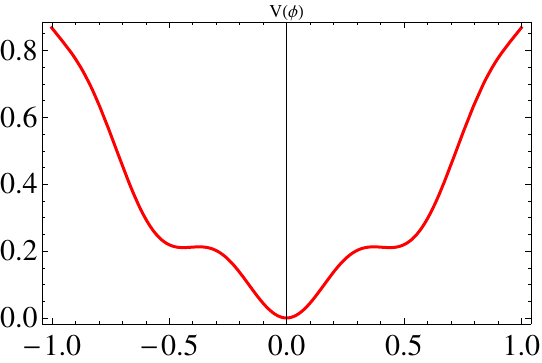}}\hspace{2cm}
	\subfigure[$(b, f, \delta, \mu)= (0.99, 0.09, 0, 0.9)$]{\includegraphics[scale=0.5]{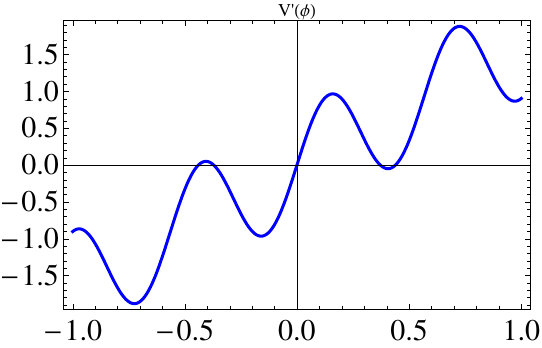} }
\end{center}\caption{\label{2PlotMonodromy-Potential} The generalized harmonic potential is $
V(\phi)= \mu ^3 \left[b f \left(\cos (\delta )-\cos \left(\delta +\frac{\phi }{f}\right)\right)+\frac{\phi ^2}{\mu}\right]$ and its derivative.} 
\end{figure}

In the figure \ref{2PlotMonodromy-Potential}, the generalized harmonic potential  $V(\phi)$ and its derivative $V^{\prime}(\phi)$ for some values of the parameters $(b, f, \delta, \mu)$ are depicted.

As in section \ref{Sect2.4}, the new variables \eqref{EQ:19} are used, they satisfy
 \begin{equation}
        3 H^2=\mu ^3 \left(\frac{f^2 v^2}{\mu }-b f \cos (\delta +v)\right)+b f \mu ^3 \cos (\delta )+\rho_c u^2.
    \end{equation}
To describe expanding universe the positive solution for $H$ of the previous equation is chosen. Hence, by introducing $\tau =\frac{\sqrt{2 \rho_c}}{f} t$, and
redefining the constants $\rho_{c}= \frac{1}{2} b f \mu ^3>0, \quad k=\frac{2 f}{b \mu}>0$, the following equations are obtained:   
\begin{align}
\label{monodronomyeqs1ab11} 
&\frac{d u}{d \tau}=-\frac{\sqrt{6}}{4} b k \mu  u \sqrt{2 \cos (\delta )+k v^2+u^2-2 \cos (\delta +v)}-k v-\sin (\delta +v), \quad \frac{d v}{d\tau}=u. 
\end{align}
The origin $(u,v)=(0,0)$ is an equilibrium point if $\delta=0$. Then, the eigenvalues of the linearization matrix of \eqref{monodronomyeqs1ab11} are 
$\left\{-\sqrt{-k-1},\sqrt{-k-1}\right\}$. The  origin is a center.

For $k\neq 0$ and $|k v_c|\leq 1$, the following are equilibrium points of \eqref{monodronomyeqs1ab11}, $(u,v)=(0,v_c)$ where $v_c$ are the roots of the transcendental equation $-k v-\sin (\delta +v)=0$. To obtain a real valued linearization matrix is additionally required that  $3 \cos (\delta )-\frac{3}{2} v_c \sin (\delta +v_c)-3 \cos (\delta +v_c)\geq 0$. The system does not admit equilibrium points $(u,v)=(0,v_c)$ other than the origin for $|k  v_c|>1$.

If $|k v_c|\leq 1$, the following are equilibrium points of \eqref{monodronomyeqs1ab11}, $(u,v)=(0,v_c)$ where $v_c$ are the roots of the transcendental equation $-k v-\sin (\delta +v)=0$. 

For $\delta\neq 0$ and $|k  v_c|\leq 1$, it is deduced that
    $\delta =-\sin ^{-1}(k v_c)-v_c$, and the  eigenvalues 
$\Big\{-\frac{1}{8} \sqrt{6 b^2 k^2 \mu ^2 \left(-2 \sqrt{1-k^2 v_c^2}+k v_c^2+2 \cos \left(\sin ^{-1}(k v_c)+v_c\right)\right)-64
   \left(\sqrt{1-k^2 v_c^2}+k\right)}$\\
	$-\frac{1}{4} b k \mu  \sqrt{-3 \sqrt{1-k^2 v_c^2}+\frac{3 k v_c^2}{2}+3 \cos \left(\sin ^{-1}(k
   v_c)+v_c\right)}$,\\
	$\frac{1}{8} \sqrt{6 b^2 k^2 \mu ^2 \left(-2 \sqrt{1-k^2 v_c^2}+k v_c^2+2 \cos \left(\sin ^{-1}(k v_c)+v_c\right)\right)-64
   \left(\sqrt{1-k^2 v_c^2}+k\right)}$\\
	$-\frac{1}{4} b k \mu  \sqrt{-3 \sqrt{1-k^2 v_c^2}+\frac{3 k v_c^2}{2}+3 \cos \left(\sin ^{-1}(k
   v_c)+v_c\right)}\Big\}$ are obtained.

For the choice of parameters $(b, f, \delta, \mu)= (0.1, 0.33, 0, 0.9)$ it follows that
   $\rho_c=\frac{24057}{2000000}\approx 0.0120285, k=\frac{22}{3}\approx 7.33333$. 
The only equilibrium point is the origin  with eigenvalues
$\left\{\frac{5 i}{\sqrt{3}},-\frac{5 i}{\sqrt{3}}\right\}$.

\begin{figure}[t!]
\begin{center}
\subfigure[\label{2monodromy-potential-1-b} $(b, f, \delta, \mu)= (0.1, 0.33, 0, 0.9)$]{\includegraphics[scale=0.48]{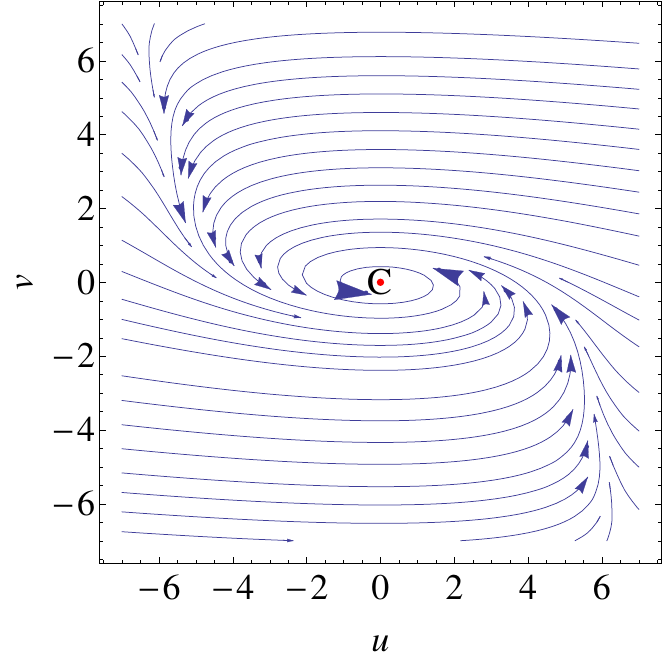}}\hspace{2cm}
\subfigure[\label{2monodromy-potential-1-a}  $(b, f, \delta, \mu)= (0.99, 0.09, 0, 0.9)$]{\includegraphics[scale=0.4]{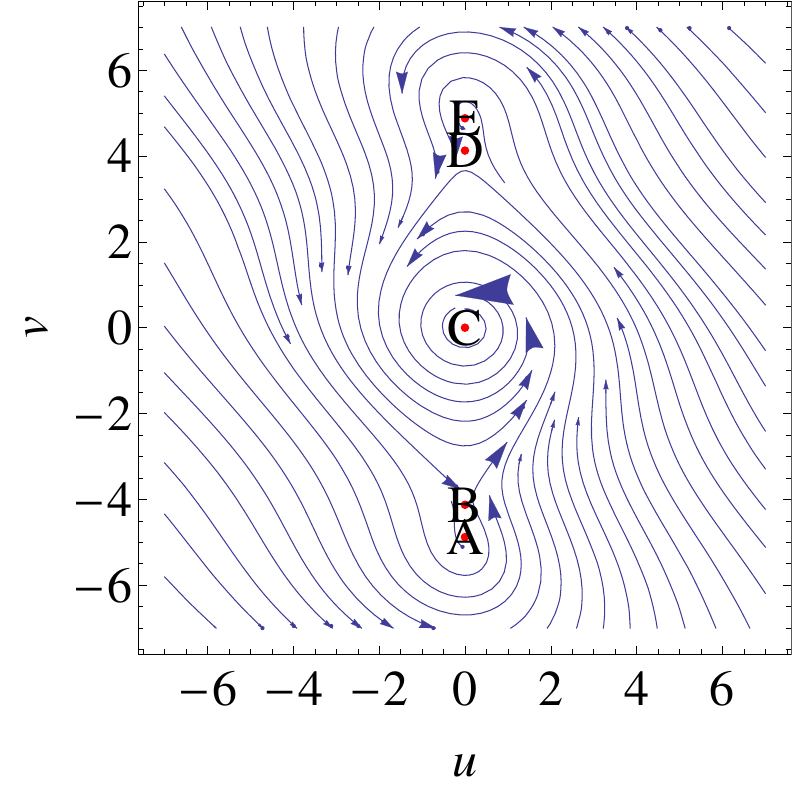}}
\caption{Phase portrait of equations    \eqref{monodronomyeqs1ab11} for some choices of parameters $(b, f, \delta, \mu)$.}
\end{center}
\end{figure}

 In figure \ref{2monodromy-potential-1-b}  some orbits of the flow of \eqref{monodronomyeqs1ab11}  are depicted for the choice of parameters $(b, 
   f, \delta, \mu)=(0.1, 0.33, 0, 0.9)$. For this choice of parameters the hypotheses and the results of Theorems \ref{tm} and \ref{thm2.1} ($\lim_{t\rightarrow \infty } \dot\phi=0$, and $\lim_{t\rightarrow \infty } \phi=0$) have been verified.  

	Setting the values $(b, 
   f, \delta, \mu)= (0.99, 0.09, 0, 0.9)$, it is obtained that
   $\rho_c=\frac{649539}{20000000}\approx 0.032477, k=\frac{20}{99}\approx 0.20202$. 
The transcendental equation is $-\frac{20 v}{99}-\sin (v)=0$. 
The equilibrium points are: 	
\begin{enumerate}
 \item $A: (u,v)=(0,-4.88035)$ with eigenvalues $-0.140267-0.591197 i, -0.140267+0.591197 i$. It is a stable spiral. 
 \item $B: (u,v)=(0,-4.12769)$ with eigenvalues $-0.749132, 0.467117$, is a saddle. 
 \item $C: (u,v)=(0,0)$ with eigenvalues $ -1.09637 i , 1.09637 i$, is a  center.
 \item $D: (u,v)=(0,4.12769)$ with eigenvalues $-0.749132, 0.467117$, is a saddle. 
 \item $E: (u,v)=(0,4.88035)$ with eigenvalues  $-0.140267-0.591197 i , -0.140267+0.591197 i$, is a stable spiral. 
\end{enumerate}
In figure \ref{2monodromy-potential-1-a}, some orbits of the flow of \eqref{monodronomyeqs1ab} are depicted and for  $(b, f, \delta, \mu)= (0.99, 0.09, 0, 0.9)$ the hypotheses of Theorem \ref{tm} hold, and the result $\lim_{t\rightarrow +\infty} \dot \phi=0$ is attained.  The hypothesis {\it{$V^{\prime}(\phi)<0$ for $\phi<0$ and $V^{\prime}(\phi)>0$ for $\phi>0$}} of Theorem \ref{thm2.1} is not verified and $\lim_{t\rightarrow +\infty}\phi$ can be zero, or finite. Recalling that this Theorem relies on the monotonicity of $V(\phi)$. Finally, the hypothesis {\it{$V^{\prime}(\phi)<0\quad \forall \phi\in\mathbb{R}$}} of Theorem \ref{thm2.2} is not fulfilled and $\lim_{t\rightarrow +\infty}\dot\phi=0, \lim_{t\rightarrow +\infty}\phi<\infty$. 
	
\subsection{Scalar field with potential  $
V(\phi)= \frac{\phi ^2}{2}+f\left[1- \cos \left(\frac{\phi }{f}\right)\right]
$, $f> 0$ non--minimally coupled to matter with coupling function $\chi(\phi)=\chi_0 e^{\frac{\lambda \phi}{4-3\gamma}}$.}
\label{SECT:New4.5}
The focus in this section is the Hubble--normalized formulation for the FLRW metric and the Bianchi I metric for a scalar field cosmology with generalized harmonic potential: 
\begin{equation}
\label{EQ:23}
 V(\phi)= \frac{\phi^2}{2} + f \left[1-\cos\left( \frac{\phi}{f}\right)\right], \quad f>0.   
\end{equation}
This is deduced by setting $\mu=\frac{\sqrt{2}}{2}, b \mu=2, \delta=0$ in equation \eqref{harmonic2}. The scalar field is non--minimally coupled to matter with coupling function  $\chi=\chi_0 e^{\frac{\lambda \phi}{4-3\gamma}}$, where  $\lambda$ is a constant and  $0\leq \gamma \leq 2, \quad \gamma \neq \frac{4}{3}$. It is assumed $\Lambda=0$. 
 For this potential the Raychaudhuri equation fails to decouple \cite{Alho:2014fha}.

The generalized harmonic potential \eqref{EQ:23}
belongs to the class of potentials studied by  \cite{Rendall:2006cq}, and  has the following generic features:
\begin{enumerate} 
    \item$V$ is a real-valued function, $V\in C^{\infty} (\mathbb{R})$, with $\lim_{\phi \rightarrow \pm \infty} V(\phi)=+\infty$. 
        \item $V$ is an even function: $V(\phi)=V(-\phi)$.
    \item  $V(\phi)$ has always a local minimum at $\phi=0$:  $V(0)=0, V'(0)=0, V''(0)>0$.
    \item There is a finite number of values $\phi_c \neq 0$ that satisfies $\phi_c +\sin \left(\frac{\phi_c}{f}\right)=0$, which are extreme points of $V(\phi)$ (local maximums or local minimums depending on $V''(\phi_c):= \frac{\cos \left(\frac{\phi_c }{f}\right)}{f}+1<0$ or $V''(\phi_c)>0$). For $\left|\phi_c\right| >1$ this set is empty. 
    \item There exist 
    $V_{max}= \max_{\phi\in [-1,1]} V(\phi)=\frac{1}{2}+f \left[1- \cos\left(\frac{1}{f}\right)\right]$,  and $V_{min}= \min_{\phi\in [-1,1]} V(\phi)=0$. The function $V$ has no upper bound, but it has a lower bound equal to zero.
  \end{enumerate}

	\begin{figure}[t]
\begin{center}
	\includegraphics[scale=0.7]{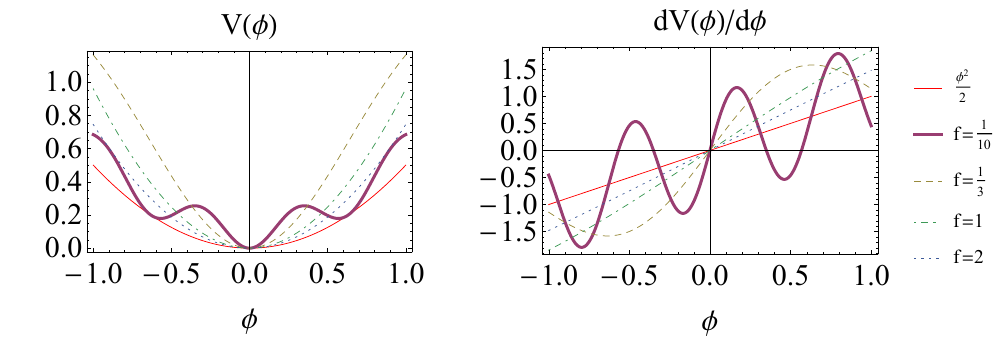}
	\caption{\label{FIG1} Generalized harmonic potential $V(\phi)=\frac{\phi ^2}{2}+f \left[1-\cos  \left(\frac{\phi }{f}\right)\right]$ (left panel) and its derivative (right panel).}
	\end{center}
\end{figure} 

In Fig. \ref{FIG1}, the generalized harmonic potential $V(\phi)=\frac{\phi ^2}{2}+ f \left[1- \cos \left(\frac{\phi }{f}\right)\right]$ and its derivative are depicted for different choices of $f$. In the limit $f\rightarrow 0$ the harmonic potential $\phi^2/2$ is recovered.

\subsubsection{Scalar field in a vacuum for flat FRW metric.}
\label{SECT:IIA}

In this section  the field equations of a scalar field with the self--interacting potential $V(\phi)$ in vacuum are studied.
Defining $y={\dot \phi}$ 
the following equations are obtained: 
\begin{subequations}
\label{eqs1}
\begin{align}
&\dot y=-\phi-\sin \left(\frac{\phi}{f}\right)-3 H y, \\
&\dot \phi=y,  \\
&\dot H= -\frac{1}{2} y^2, \label{eq44c}
\end{align}
\end{subequations}
defined on the phase space  
\begin{equation}
    \left\{(y, \phi, H)\in \mathbb{R}^3:  3 H^2=\frac{y^2}{2}+\frac{\phi^2}{2}+f \left[1- \cos\left(\frac{\phi}{f}\right)\right]\right\}.
\end{equation}
The interest is in the late time behavior of the system \eqref{eqs1} for sufficiently small positive $H$, considering that the origin is always a local minimum of $V$ with $V(0)=0$.  \newline Defining the set of local maximums of $V(\phi)$:
    \begin{equation}
    S_{+}=\left\{\phi \in \mathbb{R}: \phi +\sin \left(\frac{\phi}{f}\right)=0, V''(\phi)<0\right\}.    
    \end{equation}
If  $S_{+}= \emptyset$, the origin is the only extrema of $V(\phi)$, and it is a global minimum. Then, $V_m=V_{max}= \max_{\phi\in [-1,1]} V(\phi)$ is chosen in the following discussion. \newline
If $S_{+} \neq \emptyset$, there exists:
    \begin{equation}
        \phi_m: = \left|\underset{\phi\in S_{+}\subseteq [-1,1]}{\operatorname{arg\,min}}\, V(\phi)\right| := \{|x|:  x\in S_{+} \land \forall y \in S_{+} : V(y) \ge V(x)\}.
    \end{equation} 
Initial data such that $y(0)= y_0, \phi(0)=\phi_0 \in [-1,1], H_0=H(0),  3 H_0^2\leq V_m \leq V_{max}$, where $V_m= V (\phi_m)= \frac{\phi_m^2}{2}+f \left[1-\cos \left(\frac{\phi_m}{f}\right)\right]$ are considered. 

    	\begin{figure}[t]
\begin{center}
\subfigure[\label{FIG4a} $f=0.1$]{\includegraphics[scale=0.33]{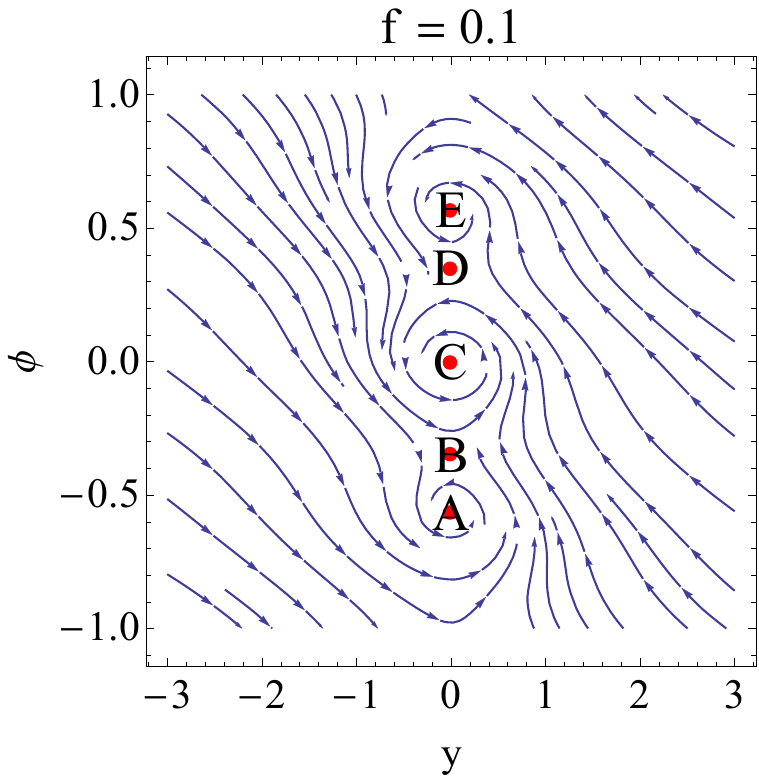}}
\subfigure[\label{FIG4b} $f=0.33$]{\includegraphics[scale=0.33]{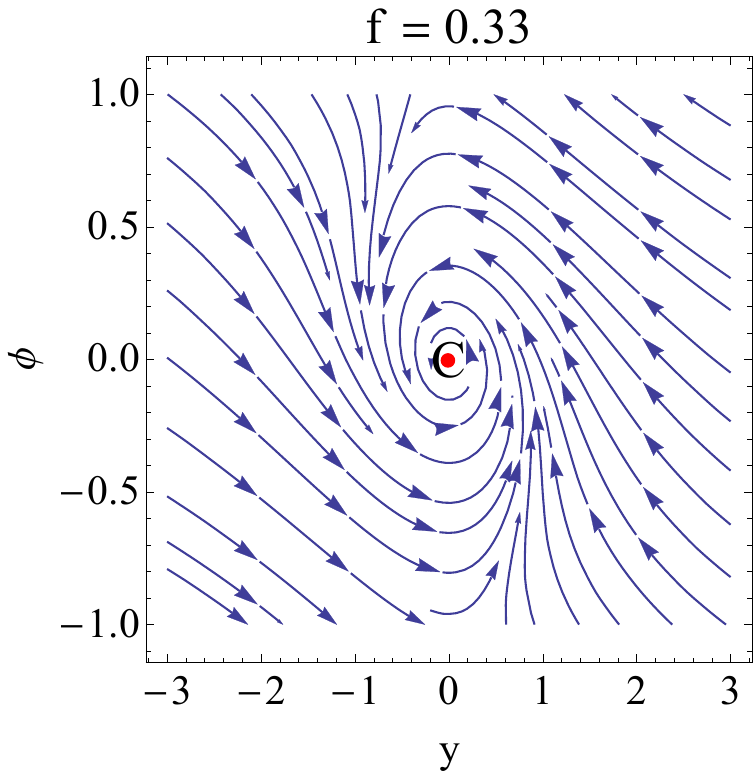}}
\subfigure[\label{FIG4c} $f=1$]{\includegraphics[scale=0.33]{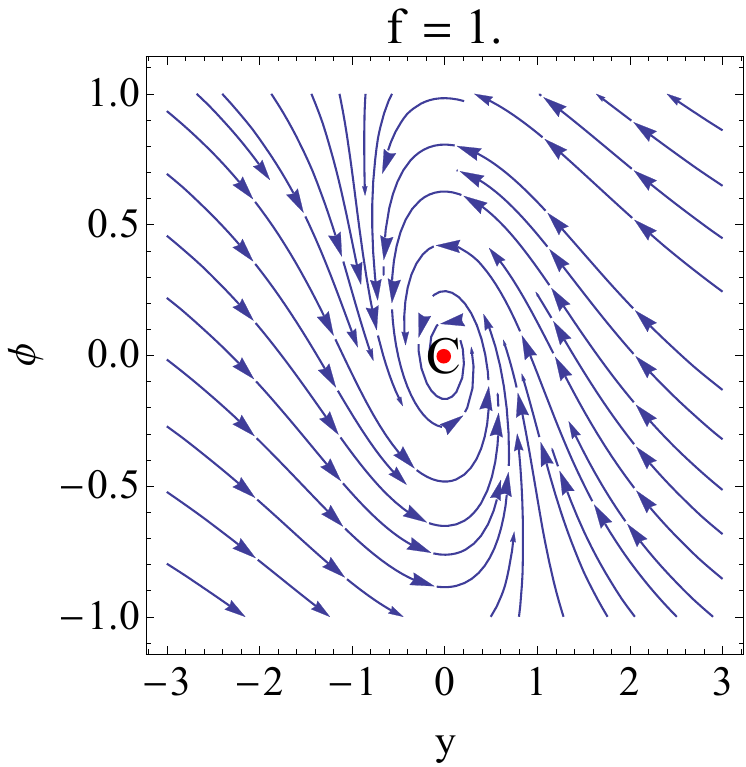}}
\caption{\label{FIG4} Phase portrait of the system \eqref{monodronomyeqs1ab} for the generalized harmonic potential $V(\phi)= \frac{\phi ^2}{2} +f\left[1- \cos \left(\frac{\phi }{f}\right)\right]$.}
\end{center}
\end{figure}

By equation \eqref{eq44c}, $H$ is monotonically decreasing, then 
\begin{equation}
\label{eq46}
3 H(t)^2:= V(\phi(t))   + \frac{1}{2} \phi(t)^2 \leq 3 H_0^2 \leq V_m, \forall t\geq 0.
\end{equation}
Hence, $V(\phi(t))\leq V_m, \forall t\geq 0$. Moreover, $-\phi_m<  \phi <\phi_m$, and there is a critical value of $y$, say $y_{\text{crit}}$, such that if $|\phi_0|<\phi_m$ and $y_0<y_{\text{crit}}$, then $|\phi(t)|$ remains less than $\phi_m$ for all $t \geq 0$. 
The equation \eqref{eq46} and the initial condition on $H(t)$ establish a maximum allowable value of $y_{\text{crit}}$,  $y_{\text{crit}}\leq \sqrt{2 V_m}$.
Therefore, if $0<H_0\leq \sqrt{V_m/3}$ and $y_0 \leq \sqrt{2 V_m}$, then, $\phi(t)$ never crosses the maximum of $V(\phi)$ throughout the evolution. Since $|\phi|$ is bounded in a neighborhood of zero  the trajectories are attracted by the origin. 

Investigating the qualitative properties of system \eqref{eqs1} can be done by analyzing the alternative equations: 
\begin{align}
\label{monodronomyeqs1ab}
&\frac{d y}{d t}=-\phi-\sin \left(\frac{\phi}{f}\right)-\sqrt{\frac{3}{2}} y \sqrt{-2 f \cos \left(\frac{\phi}{f}\right)+2 f+y^2+\phi^2}, \quad \frac{d \phi}{d t}=y,  
\end{align}
where $H$ was obtained by taking the positive branch of $3 H^2=f-f \cos \left(\frac{\phi}{f}\right)+\frac{y^2}{2}+\frac{\phi^2}{2}$.
\newline
The origin is an equilibrium point with eigenvalues $\left\{-\frac{\sqrt{-f-1}}{\sqrt{f}},\frac{\sqrt{-f-1}}{\sqrt{f}}\right\}$. It is a spiral point for $f>0$.
The system admits the equilibrium points: $(y,\phi)=(0,\phi_c)$, such that 
$\frac{\sin (\phi_c)}{f}+\phi_c=0$ with eigenvalues:\\
\begin{small}
\begin{align}
\label{Eigen3b}
&\lambda_{1,2}=-\frac{1}{2} \sqrt{\frac{3}{2}} \sqrt{\phi_c^2-2 f \left(\cos \left(\frac{\phi_c}{f}\right)-1\right)}\pm\frac{\sqrt{f \left(f \left(3 \phi_c^2+6 f-8\right)-2 \left(3 f^2+4\right) \cos \left(\frac{\phi_c}{f}\right)\right)}}{2 \sqrt{2}
   f}.
	\end{align}
\end{small}

In Fig. \ref{FIG4} the phase portrait of the equations \eqref{monodronomyeqs1ab} for the generalized harmonic potential $
V(\phi)= \frac{\phi ^2}{2} +f\left[1- \cos \left(\frac{\phi }{f}\right)\right]$ is depicted for $f=0.1$, $f=0.33$ and $f=1$.

In Fig. \ref{FIG4a} the equilibrium points  $A:(y, \phi)=(0, -0.567921)$ and $E: (y, \phi)=(0, 0.567921)$ with eigenvalues $\{-0.366359+3.01606 i,-0.366359-3.01606 i\}$ and $C: (y, \phi)=(0, 0)$ with eigenvalues $\left\{i \sqrt{11},-i \sqrt{11}\right\}$ are local attractors. They are associated to local minimums of the potential. The equilibrium points $B: (y,\phi)=(0,-0.349906)$ and $D:(y,\phi)=(0,0.349906)$ are associated with the local maximums of the potential with eigenvalues $\{-3.36281,2.48835\}$; therefore, they are saddle points. In Figs. \ref{FIG4b} and \ref{FIG4c} the origin denoted by $C$ is the unique equilibrium point of the dynamical system, corresponding to the global minimum of the potential, and it is a sink.

\subsubsection{Scalar field non--minimally coupled to matter for FLRW metric.}
In this example the field equations are
\begin{subequations}
	\label{Non_minProb2FLRW}
	\begin{align}
	&\ddot\phi+3 H \dot \phi +\phi + \sin\left( \frac{\phi}{f}\right)=\frac{\lambda}{2}\rho_m ,\\
	&\dot{\rho_m}+3\gamma H\rho_m=-\frac{\lambda}{2}\rho_m  {\dot\phi},\\
	&\dot a = a H, \\
	& \dot{H}=-\frac{1}{2}\left(\gamma \rho_m+{\dot \phi}^2\right)+\frac{k}{a^2},\\
	& 3H^2=\rho_m+\frac{1}{2}\dot\phi^2+\frac{\phi ^2}{2}+f\left[1- \cos \left(\frac{\phi }{f}\right)\right]-\frac{3 k}{a^2}.
	\end{align}
\end{subequations}

Defining the variables 
\begin{equation}
x=\frac{\dot  \phi}{\sqrt{6}H},\quad  z=\frac{1}{H},   \quad \Omega_k=-\frac{k}{a^2 H^2},
\end{equation}
and the time variable $\frac{dg}{d\tau} \equiv g'=H^{-1}\dot{g}$, the Hubble--normalized equations are the following
\begin{subequations}\label{Foster4.58}
\begin{align}
&\frac{d x}{d\tau}= -\frac{z^2 \sin \left(\frac{\phi}{f}\right)}{\sqrt{6}}+\frac{1}{2} \sqrt{\frac{3}{2}} \lambda\Omega_m-\frac{z^2 \phi}{\sqrt{6}}+3
   x^3+x \left(\frac{3 \gamma  \Omega_m}{2}+\Omega_k-3\right),\\
   &\frac{d z}{d \tau}= z \left(\frac{3 \gamma  \Omega_m}{2}+3 x^2+\Omega_k\right),\\
& \frac{d \Omega_k}{d\tau}=\Omega_k \left(3 \gamma  \Omega_m+6 x^2+2 \Omega_k-2\right), \\
&\frac{d \phi}{d \tau}= \sqrt{6} x,
\end{align}
\end{subequations}
where
\begin{equation}
    \Omega_m =1 -\Omega_k -x^2 -\frac{1}{6} \phi^2 z^2 -\frac{f}{3}z^2 \left[1-\cos\left(\frac{\phi}{f}\right)\right].
\end{equation}

The stability analysis of the equilibrium points of the system \eqref{Foster4.58} are summarized in Table \ref{Foster4.588}. Regarding the equilibrium points $P_4(\phi^*)$ the eigenvalues  $\lambda_{1,2}$ are: 
\begin{align}
& \lambda_{1,2}= -\frac{3}{2}\pm \frac{3 }{2}\sqrt{1-\frac{8 \left(\frac{1}{f}\cos \left(\frac{\phi^*}{f}\right)+1\right)}{3 \Delta}},
\end{align}
where $\phi^*$ and $f$  are related through $\sin(\phi^*/f)+\phi^*=0$, $0<|\phi^*|\leq 1$. For $z^{*}$ being real it is required that $\Delta:= {\phi^*}^2+2 f\left[1- \cos \left(\frac{{\phi^*}}{f}\right)\right]>0$.
If $\sin(\phi^*/f)+\phi^*=0$, $0<|\phi^*|\leq 1, \cos \left(\frac{\phi^*}{f}\right)+f<0$, i.e., when $\phi^*$ is a local maximum of $V(\phi)$, $P_4(\phi^*)$ is a saddle. Whenever $\sin(\phi^*/f)+\phi^*=0$, $0<|\phi^*|\leq 1, \cos \left(\frac{\phi^*}{f}\right)+f>0$, i.e., when $\phi^*$ is a non zero local minimum of $V(\phi)$, $P_4(\phi^*)$ is a sink. 
Indeed, the dynamics on the invariant manifold $x=\phi=0$ is given by 
 \begin{equation}
    \frac{d z}{d \tau}= z \left(\Omega_k-\frac{3}{2} \gamma  (\Omega_k-1)\right),\quad  \frac{d \Omega_k}{d \tau}=(3
   \gamma-2) (1-\Omega_k) \Omega_k.
 \end{equation}
 
  \begin{table}[t!]
\begin{center}
\caption{\label{Foster4.588} Stability  of the equilibrium points of \eqref{Foster4.58}. Where $z^*=\frac{\sqrt{6}}{\sqrt{{\phi^*}^2-2 f \cos \left(\frac{{\phi^*}}{f}\right)+2 f}}$.}
\resizebox{\textwidth}{!} {
\begin{tabular}{ccccc}
\hline \hline
Label & $(x, z,\Omega_k, \phi)$ & Existence & Eigenvalues & Stability\\ \hline 
$P_1(\phi_c)$ & $(0,0,0, \phi_c)$ & $\phi_c \in \mathbb{R}$ , $\lambda =0$& $\lbrace 0,\frac{3 \gamma }{2},\frac{3 \gamma }{2}-3,3 \gamma -2 \rbrace$ &  Saddle \\ 
$P_{2}(\phi_c)$ &$(0,0,1,\phi_c)$ & $\phi_c\in \mathbb{R}$ & $\lbrace -2,1,0,2-3\gamma \rbrace$ &  Saddle\\  
$P_{3}(\phi_c)$ &$(0,0,\Omega_{kc},\phi_c)$ & $\phi_c\in \mathbb{R}$, $\forall \Omega_{kc}$, $\gamma=2/3, \lambda=0$& $\lbrace -2,1,0,0\rbrace$ &  Saddle\\  
$P_4(\phi^{*})$& $(0,z^*,0,\phi^*)$& $\sin(\phi^*/f)+\phi^*=0$, $\phi^*\neq 0$ & $\lbrace -2,-3\gamma, \lambda_1,\lambda_2 \rbrace$ & See text. \\ \hline \hline
\end{tabular}}
\end{center}
\end{table}

  \begin{figure*}[t]    
\centering
     \includegraphics[scale=0.5]{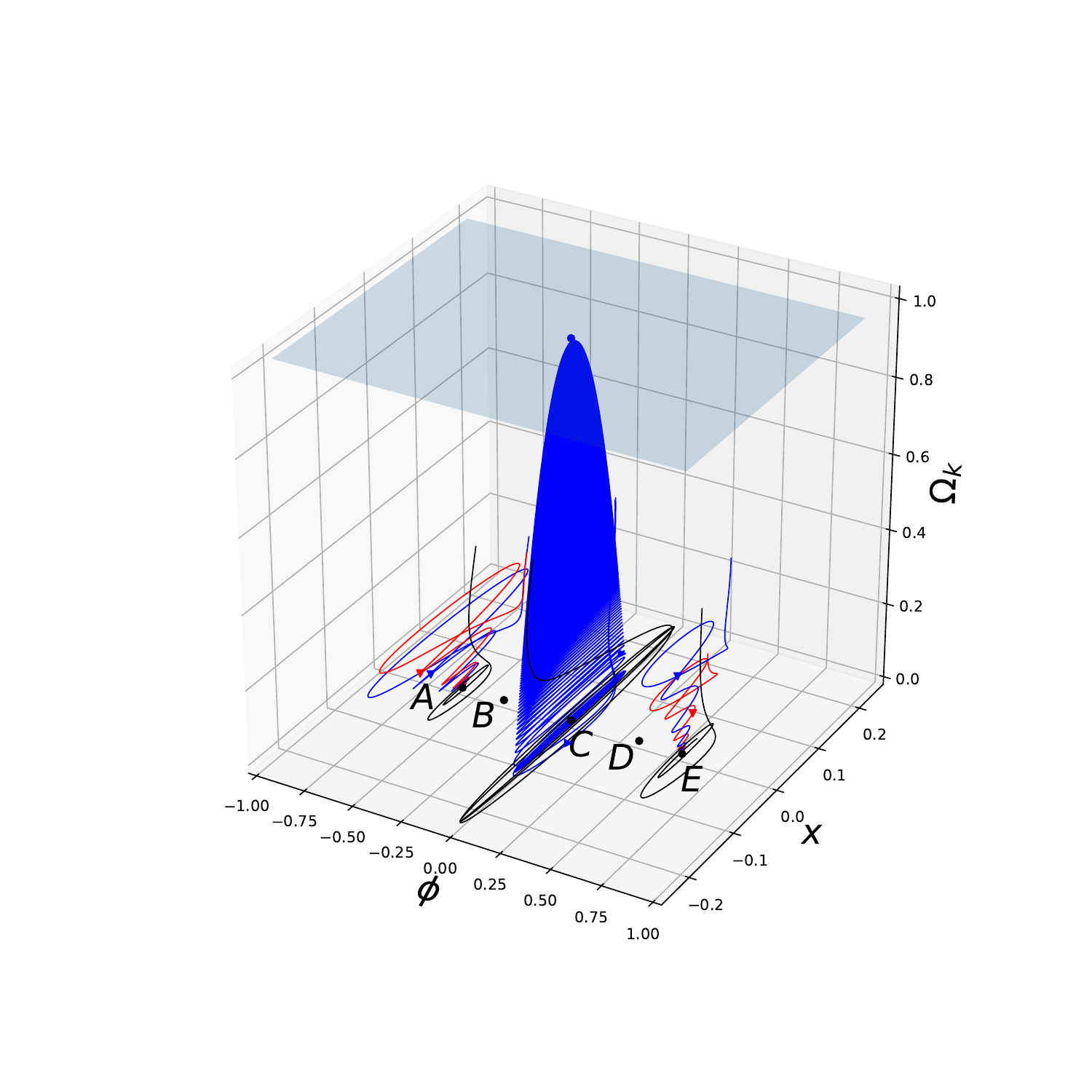}
    \caption{\label{fig:FIG7} Projections of the orbits of the system \eqref{Foster4.58} for $\gamma=1$ and $\lambda=0.1$.}
    \end{figure*}
 
       	\begin{figure}[t]
\begin{center}
\subfigure[\label{FIG8a} $\gamma=0$]{\includegraphics[scale=0.3]{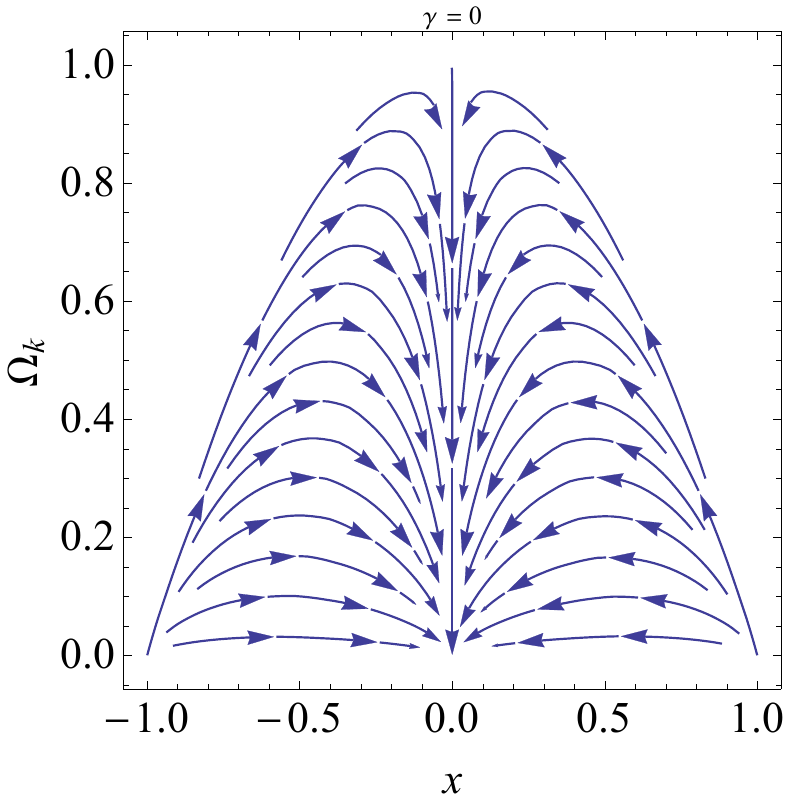}}
\subfigure[\label{FIG8b} $\gamma=2/3$]{\includegraphics[scale=0.3]{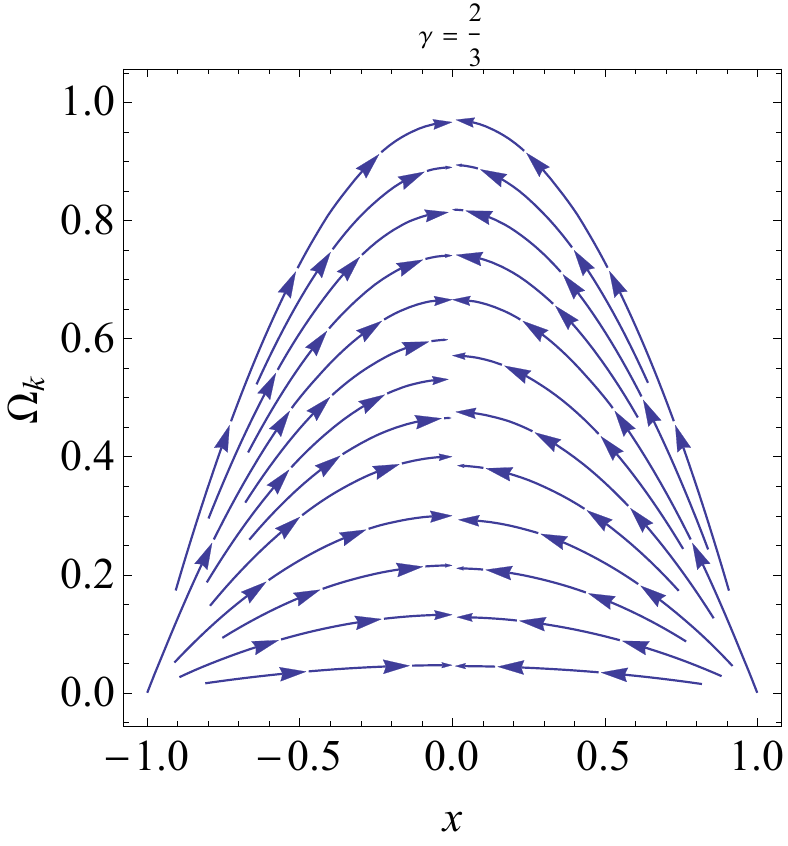}}
\subfigure[\label{FIG8c} $\gamma=1$]{\includegraphics[scale=0.3]{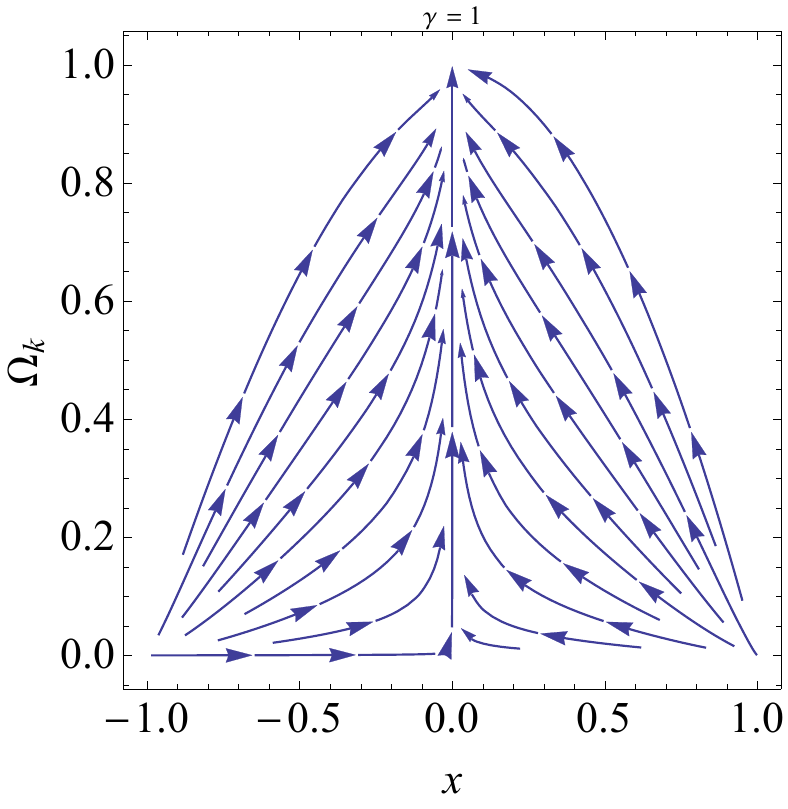}}
\caption{\label{FIG8} Phase portrait of the reduced system \eqref{syst_FIG8}.}
\end{center}
\end{figure} 
 
 The equilibrium points/lines in this invariant set are:
 \begin{enumerate}
     \item The line $z=0$ exists for $\gamma=\frac{2}{3}$. 
     \item The line $\Omega_k=0$ exists for $\gamma=0$.   
     \item The point $P_2(0)$ has coordinates $(z,\Omega_k)=(0,1)$. The eigenvalues of the reduced dynamical system are $\left\{1,-3 \gamma +2\right\}$. For $\gamma>\frac{2}{3}$, $P_2(0)$ is stable to curvature perturbations. 
     \item The point $C=P_4(0): (z,\Omega_k)=(0,0)$. The eigenvalues of the reduced dynamical system are $\left\{\frac{3 \gamma }{2},3 \gamma -2\right\}$. For $\gamma>\frac{2}{3}$, $C$  is unstable to curvature perturbations. 
 \end{enumerate}
In Figure \ref{fig:FIG7} some projections of the orbits of the system \eqref{Foster4.58} are depicted for $\gamma=1$ and $\lambda=0.1$. It is seen that the global minimum  $C$ is unstable to curvature perturbations.

As $z\rightarrow 0$ ($H \rightarrow \infty$), the following reduced system is obtained
 \begin{subequations}
  \label{syst_FIG8}
 \begin{align}
  & \frac{d x}{d \tau}=-\frac{1}{2} x \left((3 \gamma -2)\Omega_k+3 (\gamma -2)
   \left(x^2-1\right)\right),\\ 
   & \frac{d\Omega_k}{d \tau}=-\frac{1}{2} \Omega_k\left(2 (3 \gamma -2)
   (\Omega_k-1)+6 (\gamma -2) x^2\right).    
 \end{align}
 \end{subequations}
 The dynamics on the invariant set $z=0$, given by the reduced system \eqref{syst_FIG8}, is presented in Figure \ref{FIG8}.
  Therefore,  the result in \cite{Giambo:2019ymx} is verified, that is for $\gamma> 2/3$,  in a non-degenerated minima with zero critical value the curvature has a dominant effect on the late evolution of the universe and will eventually dominates both the perfect fluid and the scalar field.

\subsubsection{Scalar field non--minimally coupled to matter for Bianchi I metric.}

The equations of motion are:  
\begin{subequations}
	\label{Non_minProb2BI}
	\begin{align}
	&\ddot\phi+3 H \dot \phi +\phi + \sin\left( \frac{\phi}{f}\right)=\frac{\lambda}{2}\rho_m ,\\
	&\dot{\rho_m}+3\gamma H\rho_m=-\frac{\lambda}{2}\rho_m  {\dot\phi},\\
	&\dot a = a H, \\
	& \dot{H}=-\frac{1}{2}\left(\gamma \rho_m+{\dot \phi}^2\right)-\frac{\sigma_0^2}{a^6},\\
	& 3H^2=\rho_m+\frac{1}{2}\dot\phi^2+\frac{\phi ^2}{2}+f\left[1- \cos \left(\frac{\phi }{f}\right)\right]+\frac{\sigma_0^2}{a^6}.
	\end{align}
\end{subequations}

\begin{table}[t!]
\begin{center}
\caption{\label{Foster4.4744} Stability of the equilibrium points of system \eqref{Foster4.74}. Where $z^*=\frac{\sqrt{6}}{\sqrt{{\phi^*}^2-2 f \cos \left(\frac{\phi^*}{f}\right)+2 f}}$.}
\resizebox{\textwidth}{!}{
\begin{tabular}{ccccc}
\hline\hline
Label & $(x,z,\Sigma, \phi)$ & Existence  & Eigenvalues & Stability\\  \hline
$Q_1(\phi_c)$ & $(0,0,0,\phi_c)$ & $\phi_c \in \mathbb{R}$, $\lambda=0$ & $\lbrace 0,\frac{3 \gamma }{2},\frac{3 (\gamma -2)}{2},\frac{3 (\gamma -2)}{2}\rbrace$ &  Saddle \\ 
$Q_{2}(\phi_c)$ &$(0,0,-\sqrt{3}, \phi_c)$ & $\phi_c \in \mathbb{R}$ & $\lbrace 0,0,3, 3(2-\gamma) \rbrace$ & 2D unstable manifold\\ 
$Q_{3}(\phi_c)$ &$(0,0,\sqrt{3}, \phi_c)$ & $\phi_c\in \mathbb{R}$ & $\lbrace 0,0,3, 3(2-\gamma) \rbrace$ &  2D unstable manifold \\
$Q_4(\phi^{*})$& $(0,z^*,0,\phi^*)$& $\sin(\phi^*/f)+\phi^*=0$ & $\lbrace -3,-3\gamma, \lambda_1,\lambda_2 \rbrace$ & See text. \\ \hline\hline 
\end{tabular}}
\end{center}
\end{table}
\begin{figure*}[t]    
\centering
     \includegraphics[scale=0.5]{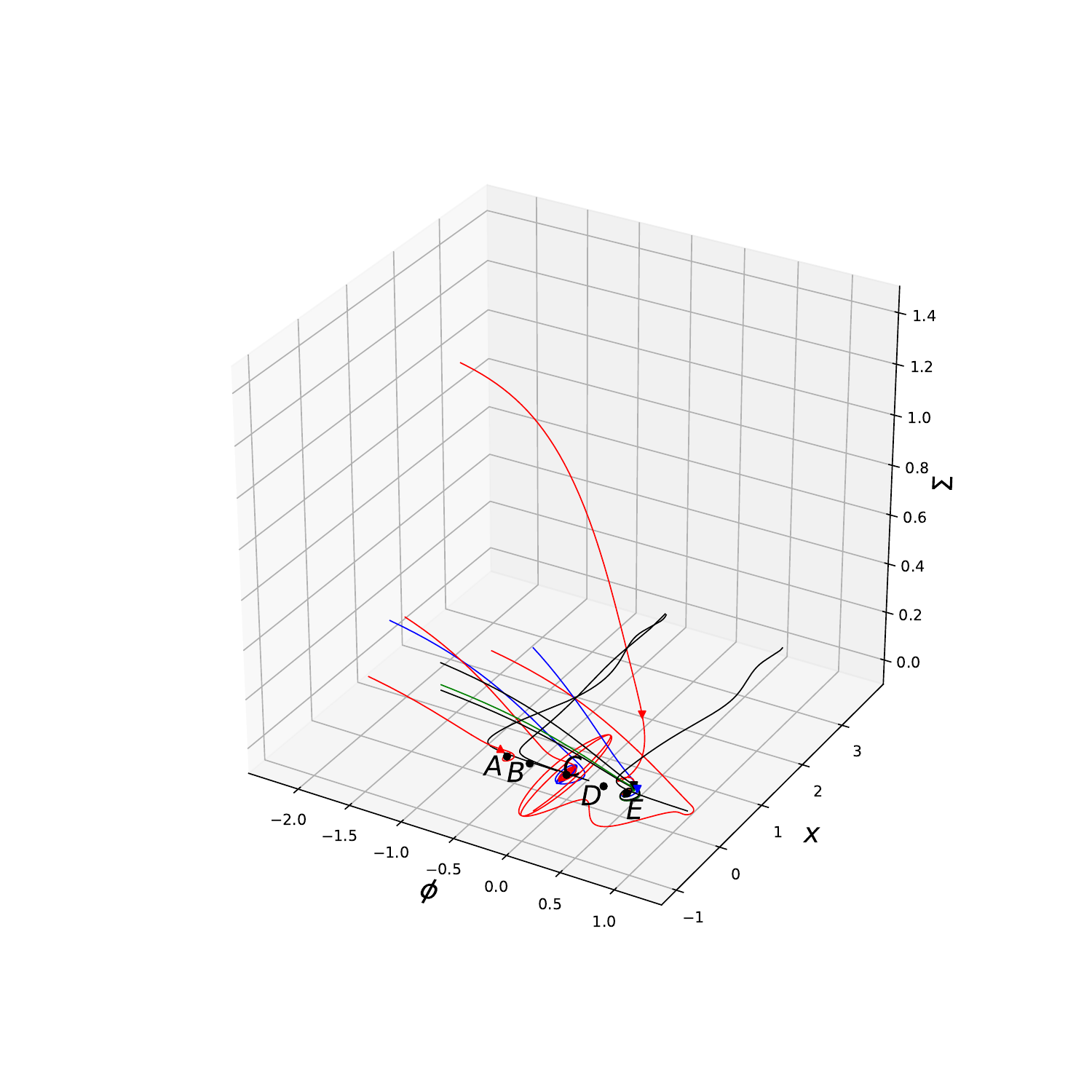}
     \hspace{-0.5cm}
    \caption{\label{fig:FIG7D} Projections of the orbits of the system \eqref{Foster4.74} for $\gamma=1$ and $\lambda=0.1$.}
    \end{figure*}
Defining the variables: 
\begin{equation}
x=\frac{\dot  \phi}{\sqrt{6}H}, \quad z=\frac{1}{H}, \quad \Sigma=\frac{\sigma_0}{H^2 a^3},
\end{equation}
and the time variable   $\frac{dg}{d\tau} \equiv g'=H^{-1}\dot{g}$, the Hubble--normalized equations are the following:
\begin{subequations}\label{Foster4.74}
\begin{align}
&\frac{d z}{d \tau}= z \left(\frac{3 \gamma  \Omega_m}{2}+\Sigma ^2+3 x^2\right),\\
&\frac{d x}{d \tau}= -\frac{z^2 \sin \left(\frac{\phi}{f}\right)}{\sqrt{6}}+\frac{1}{2} \sqrt{\frac{3}{2}} \lambda  \Omega_m-\frac{z^2 \phi}{\sqrt{6}}+3 x^3+x\left(\frac{3 \gamma  \Omega_m}{2}+\Sigma ^2-3\right),\\
&\frac{d \Sigma}{d \tau}= \Sigma ^3+\frac{3}{2} \Sigma  \left(\gamma  \Omega_m+2 x^2-2\right),\\
&\frac{d \phi}{d \tau}= \sqrt{6} x,
\end{align}
\end{subequations}
where
\begin{equation}
    \Omega_m =1 -\frac{1}{3}\Sigma^2 -x^2 -\frac{1}{6}\phi^2 z^2 -\frac{f}{3}z^2 \left[1-\cos\left(\frac{\phi}{f}\right)\right].
\end{equation}

  \begin{figure}[t]
\begin{center}
\subfigure[\label{FIG9a} $\gamma=0$]{\includegraphics[scale=0.33]{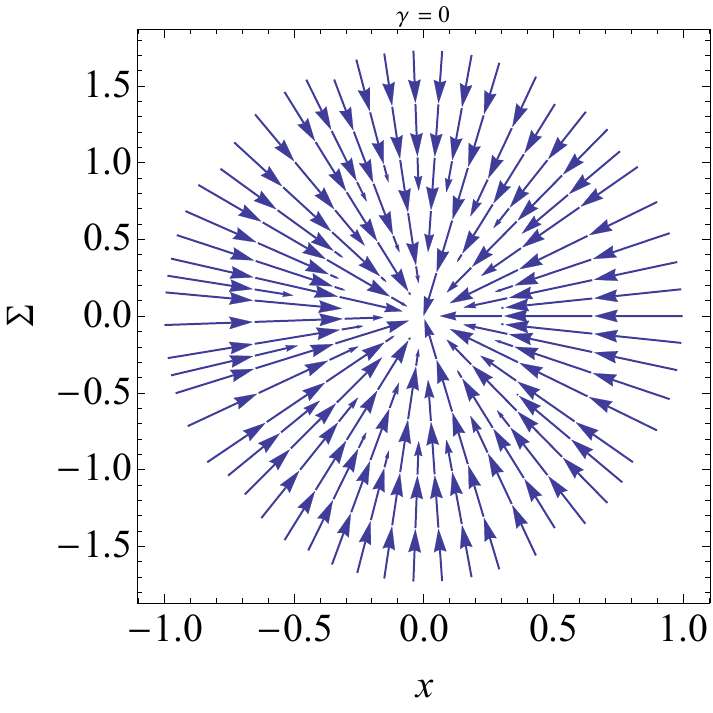}}
\subfigure[\label{FIG9b} $\gamma=2/3$]{\includegraphics[scale=0.33]{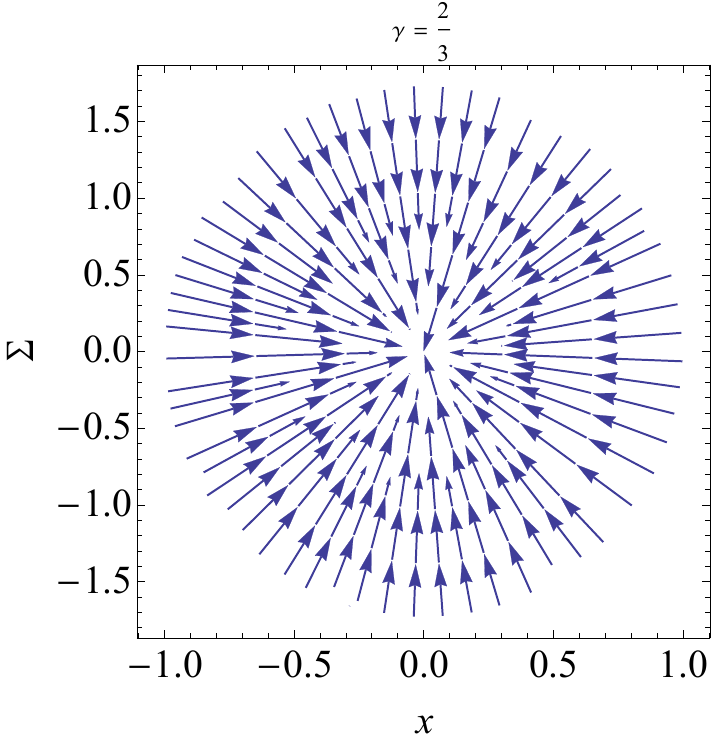}}
\subfigure[\label{FIG9c} $\gamma=1$]{\includegraphics[scale=0.33]{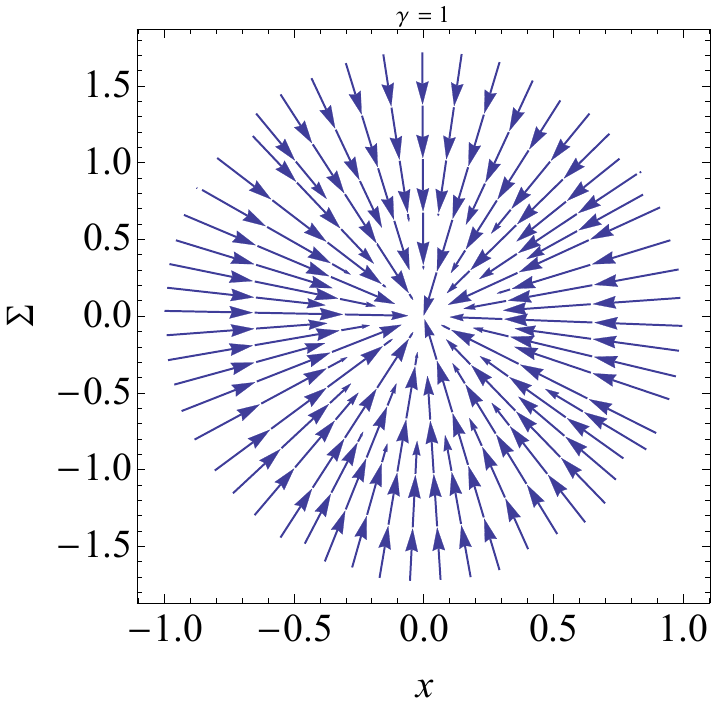}}
\caption{\label{FIG9} Phase portrait of the reduced system \eqref{syst_FIG9}.}
\end{center}
\end{figure}

The equilibrium points of the system \eqref{Foster4.74}  and their existence and stability conditions are summarized in Table \ref{Foster4.4744}.

 Regarding the equilibrium points $Q_4(\phi^*)$ the eigenvalues  $\lambda_{1,2}$ are: 
\begin{align}
& \lambda_{1,2}= -\frac{3}{2}\pm\frac{3 }{2}\sqrt{1-\frac{8 \left(\frac{1}{f}\cos \left(\frac{\phi^*}{f}\right)+1\right)}{3 \Delta}},
\end{align}
where $\phi^*$ and $f$  are related through $\sin(\phi^*/f)+\phi^*=0$, $|\phi^*|\leq 1$. For $z^{*}$ being real it is required that $\Delta:= {\phi^*}^2+2 f\left[1- \cos \left(\frac{{\phi^*}}{f}\right)\right]>0$.
If $\sin(\phi^*/f)+\phi^*=0$, $|\phi^*|\leq 1, \cos \left(\frac{\phi^*}{f}\right)+f<0$, i.e., when $\phi^*$ is a local maximum of $V(\phi)$, $Q_4(\phi^*)$ is a saddle. Whenever $\sin(\phi^*/f)+\phi^*=0$, $|\phi^*|\leq 1, \cos \left(\frac{\phi^*}{f}\right)+f>0$, i.e., when $\phi^*$ is a local  minimum of $V(\phi)$, $Q_4(\phi^*)$ is a sink. In Figure \ref{fig:FIG7D} some projections of the orbits of the system \eqref{Foster4.74} are depicted for $\gamma=1$ and $\lambda=0.1$.

 The dynamics on the invariant set $x=\phi=0$ is given by
\begin{equation}
   \frac{d z}{d \tau}= z \left(\Sigma ^2-\frac{1}{2} \gamma  \left(\Sigma ^2-3\right)\right),\quad \frac{d\Sigma}{d \tau}=-\frac{1}{2} (2-\gamma ) \Sigma  \left(3-\Sigma ^2\right).
\end{equation}
 The equilibrium points/lines in this invariant set are:
 \begin{enumerate}
     \item The line $z=0$ exists for $\gamma=2$. 
     \item The line $\Sigma=0$ exists for $\gamma=0$.   
     \item The point  $Q_{2,3}(0): (z,\Sigma)=(0,\mp\sqrt{3})$. The eigenvalues of the reduced dynamical system are $\{3,3 (2-\gamma )\}$. They are local sources in the invariant set. 
     \item  The point  $Q_4(0): (z,\Sigma)=(0,0)$ with eigenvalues $\left\{\frac{3 (\gamma -2)}{2},\frac{3 \gamma }{2}\right\}$ is a saddle (unstable to shear perturbations).
 \end{enumerate}

As $z\rightarrow 0$ ($H \rightarrow \infty$), we obtain the reduced system
 \begin{equation}
  \label{syst_FIG9}
   \frac{d x}{d \tau}=\frac{1}{2} (2-\gamma) x \left(\Sigma ^2+3 x^2-3\right),\quad \frac{d\Sigma}{d \tau}= \frac{1}{2} (2-\gamma) \Sigma  \left(\Sigma ^2+3 x^2-3\right).
\end{equation}
The physical solutions with $\Sigma ^2+3 x^2\leq 3$ tends radially to the origin, as it is shown in Figure \ref{FIG9}.

\section{Discussion}
\label{discussion}
 In this research   some theorems were presented and the  asymptotic behavior of a very general cosmological model consisting of a scalar field non--minimally coupled to matter was analyzed, where  a ``geometric'' term $G_0(a)$ is included. This term mainly represents the spatial curvature in FLRW models or the anisotropy in Bianchi I metrics. However, the proofs of the theorems have been made in full generality to incorporate the inverse power-law  ``geometric'' term $G_0(a)$; such that, it can effectively behave like a radiation fluid (see, e.g., \cite{Fadragas:2014mra}), wherein the energy density decays as $\propto a^{-4}$ or as a stiff fluid wherein the energy density decays as $\propto a^{-6}$.  Any effective non-negative energy density which depends on the scale factor $a$ may be considered as sub-cases of the present model as well. The coupling function is proposed such that the scalar field formulation of $f(R)$- gravity is incorporated as well as a particular case of our scenario. 
New Theorems: \ref{Proposition I}, \ref{Theorem5.2}, \ref{Proposition II}, \ref{Theorem5.3}, \ref{Proposition III} and \ref{thmIIIFINAL} valid for general situations in the context of scalar field cosmologies with arbitrary potential and / or with arbitrary couplings to matter were proven. Some well-known results from the literature are recovered, and they are presented as the corollaries \ref{PropositionIb}, \ref{tm}, \ref{PropositionII}, \ref{thm2.1}, \ref{Prop4Miritzis} and \ref{thm2.2}.
Both local and global dynamical system variables and smooth transformations of the scalar field were used to provide qualitative features of the model at hand. Conditions of a scalar field potential $V\in  C^{2}(\mathbb{R})$  under which $\displaystyle{\lim_{t\rightarrow  \infty} \dot \phi =0}$  were discussed.  These conditions are very general: non-negativity of the potential which is zero only on the origin and the boundedness of both  $V^{\prime}(\phi)$ and  $V(\phi)$ (Theorem \ref{tm}). Additionally,  some extra conditions for having $\displaystyle{\lim_{t\rightarrow  \infty} \phi(t) \in \lbrace -\infty , 0 , + \infty \rbrace }$ were presented. They are the previous conditions with the addition of $V^{\prime}(\phi)>0$ for $\phi>0$ and $V^{\prime}(\phi)<0$ for $\phi<0$ (Theorem \ref{thm2.1}). 
Mild conditions under the potential (satisfied by the exponential potential with negative slope) for having  $\lim_{t \rightarrow  + \infty} \dot \phi =0$ and $\lim_{t \rightarrow  + \infty} \phi (t)= +\infty$ (Theorem \ref{thm2.2}) were also considered. 
The flexibility of the hypotheses of the Theorems was explored in order to obtain the same conclusions or provide a counterexample.  In particular, cosine-like corrections with a small phase: $V_1(\phi)= \mu^3 \left[\frac{\phi^2}{\mu} + b f \cos\left(\delta + \frac{\phi}{f}\right)\right]$, $b\neq 0$ and 
$V_2(\phi)= \mu ^3 \left[b f \left(\cos (\delta )-\cos \left(\delta +\frac{\phi }{f}\right)\right)+\frac{\phi ^2}{\mu}\right]
$, $b\neq 0$  were incorporated. The motivation for this kind of potential's correction is found in the context of inflation in loop- quantum cosmology \cite{Sharma:2018vnv}.
In Section \ref{Sect2.4} a qualitative analysis for a scalar-field cosmology with generalized harmonic potential $V_1(\phi)$ was presented, whereas in Section \ref{Sect.2.5} a qualitative analysis for a scalar-field cosmology with generalized harmonic potential $V_2(\phi)$ was presented.

In the first generalization of the harmonic potential $V_1(\phi)$, some instances which verified the hypothesis and the results of Theorems \ref{tm}, \ref{thm2.1} ($\lim_{t\rightarrow \infty } \dot\phi=0$ and $\lim_{t\rightarrow \infty } \phi=0$) were found; as well as situations when the hypotheses {\it{$V(\phi)\geq 0$ and $V(\phi)=0$, if and only if $\phi=0$}} of Theorem \ref{tm} are not satisfied, though the result $\lim_{t\rightarrow +\infty} \dot \phi =0$ still holds. The hypotheses {\it{$V(\phi)\geq 0$ and $V(\phi)=0$ if and only if, $\phi=0$}} and {\it{$V^{\prime}(\phi)<0$ for $\phi<0$ and $V^{\prime}(\phi)>0$ for $\phi>0$}} of Theorem \ref{thm2.1} are not satisfied, and $\lim_{t\rightarrow +\infty}\phi$ can be finite (rather than zero or infinity).  When the hypotheses {\it{$V(\phi)\geq 0$, and $V^{\prime}(\phi)<0\quad \forall \phi\in\mathbb{R}$}} of Theorem \ref{thm2.2} are not fulfilled and $\lim_{t\rightarrow +\infty}\dot\phi=0, \lim_{t\rightarrow +\infty}\phi<\infty$. In addition,  some instances for the potential $V_2(\phi)$ where the hypotheses and the results of the Theorems \ref{tm} and \ref{thm2.1} are verified were found, since ($\lim_{t\rightarrow \infty } \dot\phi=0$, and $\lim_{t\rightarrow \infty } \phi=0$).  For other choices of the parameters the hypothesis of Theorem \ref{tm} holds, and the result $\lim_{t\rightarrow +\infty} \dot \phi=0$ is attained.  The hypothesis {\it{$V^{\prime}(\phi)<0$ for $\phi<0$ and $V^{\prime}(\phi)>0$ for $\phi>0$}} of Theorem \ref{thm2.1} are not satisfied and $\lim_{t\rightarrow +\infty}\phi$ can be zero or finite. Recalling this Theorem relies on the monotonicity of $V(\phi)$. The hypotheses {\it{$V^{\prime}(\phi)<0\quad \forall \phi\in\mathbb{R}$}} of Theorem \ref{thm2.2} are not fulfilled and $\lim_{t\rightarrow +\infty}\dot\phi=0, \lim_{t\rightarrow +\infty}\phi<\infty$.  In other words,  some simple examples that do not satisfy one or more  hypotheses of the Theorems proved were discussed, obtaining some counterexamples. 

Finally, in section \ref{SECT:New4.5} the Hubble--normalized formulation for FLRW metric and for the Bianchi I metric for a scalar field cosmology with the generalized harmonic potential $V(\phi)= \frac{\phi^2}{2} + f \left[1-\cos\left( \frac{\phi}{f}\right)\right]$, $f> 0$, non--minimally coupled to matter with coupling function  $\chi=\chi_0 e^{\frac{\lambda \phi}{4-3\gamma}}$, where  $\lambda$ is a constant and  $0\leq \gamma \leq 2, \quad \gamma \neq \frac{4}{3}$ was used. The late time attractors are associated to equilibrium points  $\left(\frac{\dot \phi}{\sqrt{6}H},\frac{1}{H}, \frac{G_0(a)}{3 H^2}, \phi\right)=\left(0,\frac{\sqrt{6}}{\sqrt{{\phi^*}^2-2 f \cos \left(\frac{\phi^*}{f}\right)+2 f}},0,\phi^*\right)$,  whenever $\sin(\phi^*/f)+\phi^*=0$, $0<|\phi^*|\leq 1, \cos \left(\frac{\phi^*}{f}\right)+f>0$, i.e., when $\phi^*$ is a local non zero minimum of $V(\phi)$. That is, the conclusion of Theorem \ref{thmIIIFINAL} is achieved. For FLRW metrics, the global minimum  $C=P_4(0): (z,\Omega_k)=(0,0)$ is unstable to curvature perturbations for $\gamma>\frac{2}{3}$. Therefore, the result in \cite{Giambo:2019ymx} is confirmed, that for $\gamma> 2/3$,  in a non-degenerated minima with zero critical value the curvature has a dominant effect on the late evolution of the universe, and will eventually dominates both the perfect fluid and the scalar field. For Bianchi I model, the global minimum $Q_4(0): (z,\Sigma)=(0,0)$, with $V(0)=0$,  is unstable to shear perturbations.

\section{Conclusions}
\label{Sect:7}

In this paper, phase-space descriptions  of generalized scalar field cosmologies with arbitrary potentials and arbitrary couplings to matter were used to find qualitative features of solutions. New Theorems were proved and  previous results that were recovered as Corollaries of the present results were also retrieved. Examples and counterexamples of the Theorems were presented, considering scalar field's potentials with small cosine-like corrections motivated by loop-quantum cosmology. Finally, the Hubble--normalized formulation for FLRW metric and for the Bianchi I metric for a scalar field cosmology with generalized harmonic potential, non--minimally coupled to matter was used.  The main difficulty is that a transcendental equation needs to be solved.  Additionally, the Hubble normalized equations are augmented by the Raychaudhuri equation, and the resulting system is difficult to analyze with the usual dynamical systems approach. Other approaches, such as a local stability analysis, are also difficult to apply due to the oscillations entering the system via Klein--Gordon equation \cite{Fajman:2020yjb}. Complementary formulations based on  \cite{Alho:2014fha} are implemented in a companion paper  \cite{PaperII}.

\ack

This research was funded by  Agencia Nacional de Investigaci\'on y Desarrollo - ANID through the program FONDECYT Iniciaci\'on grant no.
11180126 and by Vicerrector\'{\i}a de Investigación y Desarrollo Tecnol\'ogico at
Universidad Cat\'olica del Norte. Ellen de los Milagros Fern\'andez Flores, Alfredo Millano, and Joey Latta are acknowledged for proofreading this manuscript and improving the English. Sebasti\'an Cu\'ellar is acknowledged for producing  Figures \ref{fig:FIG7} and \ref{fig:FIG7D}. Thanks to Alan Coley for his encouraging suggestions.

\end{document}